\documentclass[aps,pre,twocolumn,showpacs,superscriptaddress,amsmath,amssymb]{revtex4-1}
\usepackage{amsmath}

\usepackage[utf8]{inputenc}
\usepackage[portuguese,english]{babel}

\usepackage{graphicx}% Include figure files
\usepackage{dcolumn}% Align table columns on decimal point
\usepackage{bm}% bold math
\usepackage{xcolor}
\usepackage{cases} %For subnumcases
\usepackage{ulem}

\begin{document}

\title{Smoluchowski equations for linker-mediated irreversible aggregation}

  \author{J. M. Tavares}
   \email{jmtavares@fc.ul.pt}
    \affiliation{Centro de F\'{i}sica Te\'{o}rica e Computacional, Universidade de Lisboa, 1749-016 Lisboa, Portugal}
    \affiliation{Instituto Superior de Engenharia de Lisboa, ISEL, Avenida Conselheiro Em\'{i}dio Navarro, 1  1950-062 Lisboa, Portugal}

  \author{G. C. Antunes}
   \email{antunes@is.mpg.de}
    \affiliation{Departamento de F\'{\i}sica, Faculdade de Ci\^{e}ncias, Universidade de Lisboa, 
    1749-016 Lisboa, Portugal}
   \affiliation{Centro de F\'{i}sica Te\'{o}rica e Computacional, Universidade de Lisboa, 1749-016 Lisboa, Portugal}
    \affiliation{Max Planck Institute for Intelligent Systems, Stuttgart, Germany}
    \affiliation{Institute for Theoretical Physics IV, University of Stuttgart, Pfaffenwaldring 57, 70569 Stuttgart, Germany}

  \author{C. S. Dias}
   \email{csdias@fc.ul.pt}
    \affiliation{Departamento de F\'{\i}sica, Faculdade de Ci\^{e}ncias, Universidade de Lisboa,     1749-016 Lisboa, Portugal}
    \affiliation{Centro de F\'{i}sica Te\'{o}rica e Computacional, Universidade de Lisboa, 1749-016 Lisboa, Portugal}

  \author{M. M. Telo da Gama}
   \email{mmgama@fc.ul.pt}
    \affiliation{Departamento de F\'{\i}sica, Faculdade de Ci\^{e}ncias, Universidade de Lisboa,     1749-016 Lisboa, Portugal}
    \affiliation{Centro de F\'{i}sica Te\'{o}rica e Computacional, Universidade de Lisboa, 1749-016 Lisboa, Portugal}

  \author{N. A. M. Ara\'ujo}
   \email{nmaraujo@fc.ul.pt}
    \affiliation{Departamento de F\'{\i}sica, Faculdade de Ci\^{e}ncias, Universidade de Lisboa,    1749-016 Lisboa, Portugal}
    \affiliation{Centro de F\'{i}sica Te\'{o}rica e Computacional, Universidade de Lisboa, 1749-016 Lisboa, Portugal}

\begin{abstract}
We developed a generalized Smoluchowski 
 framework to study linker-mediated aggregation,
where linkers and particles are explicitly taken into account. We assume that the bonds between linkers and particles are irreversible, and that clustering occurs through limited diffusion aggregation. The kernel is chosen by analogy with single-component diffusive aggregation but the clusters are distinguished by their number of particles and linkers. 
%in each cluster. 
We found that the dynamics depends on three relevant factors, all tunable experimentally: (i) the ratio of the diffusion coefficients of particles and linkers; (ii) the relative number of particles and linkers; and (iii) the maximum number of linkers that may bond to a single particle. 
%We solve these equations by simulating a lattice model of %linkers and
%particles whose time evolution is determined by the %generalized Smoluchowski equation and analytically using a %scaling hypothesis for the same equation. 
To solve the Smoluchoski equations analytically we employ a scaling hypothesis that renders the fraction of bondable sites of a cluster independent of the size of the cluster, at each instant. We perform numerical simulations of the corresponding lattice model to test this hypothesis. 
We obtain results for the asymptotic limit, and the time evolution of the bonding probabilities and the size distribution of the clusters. These findings are in agreement with experimental results reported in the literature and shed light on unexplained experimental observations.

\end{abstract}

\maketitle

\section{Introduction \label{sec.intro}}

The self-assembly of micro- amd nano-size particles mediated by linkers has been the focus of extensive experimental
\cite{Hiddessen2000,Milam2003,Hiddessen2004,Ghofraniha2009,Bharti2014,Peng2016,Singh2015,Castanon2018,Lowensohn2019}, 
theoretical \cite{Antunes2019,Sciortino2009,Lindquist2016}
and simulation \cite{Ghofraniha2009,Antunes2019,Sciortino2009,Corezzi2010,Cyron2013,Pierce2004,Wang2020a,Xia2020}
studies in recent years, since the addition of a second species provides extra control of the aggregation process. Indeed the linkers to particles ratio
\cite{Hiddessen2000,Ghofraniha2009,Peng2016,Castanon2018,Lowensohn2019},  
the strength of the linker-particle interactions
\cite{Bharti2014,Singh2015}, 
and the maximum number of linkers that can bond to one particle \cite{Castanon2018,Singh2015}, 
may be  used to tune the stability of the mixture,
 the extent and topology of the assembled structures,
  the equilibrium phase behavior and the rheological properties of the system \cite{Hiddessen2004,Muller2014}. 
 The linker-particle interactions are most often engineered using complementary biological macromolecules, 
 such as complementary DNA strands \cite{Hiddessen2004,Castanon2018} or streptavidin and biotin \cite{Hiddessen2000,Milam2003,Ghofraniha2009}. As a consequence, the bonds formed between particles mediated by linkers are extremely strong. These systems form several types of disordered and more or less open clusters, even at extremely low volume fractions \cite{Hiddessen2004,Ghofraniha2009,Castanon2018}. 
Theoretical studies have focused mostly on equilibrium properties~\cite{Cyron2013,Lindquist2016}. Some simulations have addressed the dynamics of these systems, either to get information about the early stages of aggregation~\cite{Ghofraniha2009} or to try 
to predict the type of assembled structures at long times~\cite{Pierce2004,Wang2020a}.  An exception is the work reported in \cite{Sciortino2009}, where a theory for the 
dynamics of aggregation of linker-particle binary mixtures was developed and studied.
%No theoretical description of the aggregation dynamics that includes explicitly the effect of the linkers  has been attempted, to the best of our %knowledge.

The simplest theoretical description of the kinetics of aggregation is based on the Smoluchowski coagulation equation 
\cite{Chandrasekhar43,Krapivsky2011}: two clusters of mass $i$ and $j$  merge at rate 
$K_{ij}$ to form a cluster of mass $i+j$. This description is general and flexible \cite{VanDongen1984,Aldous1999,Leyvraz1983}: its application to a particular type of aggregation is encoded in $K_{ij}$ - the so called kernel.  The choice of the appropriate kernel for a given aggregation process results from a wise combination of (intuitive) knowledge of the specificities of the clustering physical mechanism and the ability to draw relevant information from the solution of the resulting equation \cite{Smit1994}. 
The wide variety of kernels that have been studied \cite{Aldous1999,Leyvraz1983} depend on the size or mass of the aggregating clusters, but not on the presence of other components that promote aggregation. This includes the only proposal (to the best of our knowledge) to use Smoluchoswki equations to study linker particle aggregation \cite{Sciortino2009}: the effect of linkers was subsumed into an effective functionality and a single time-scale for bonding was considered. These assumptions led to the use of the polymerization kernel \cite{VanDongen1984} that is appropriate for single particle aggregation. 
 
The goal of this work is to develop a generalized Smoluchowski equation for linker-mediated aggregation where linkers and particles are explicitly taken into account. 
 Inspired by the experimental works of \cite{Hiddessen2000,Ghofraniha2009,Castanon2018}, we will assume that the bonds between linkers and particles are irreversible and that clustering happens when there is an encounter between diffusive linkers and particles (limited diffusion aggregation). The choice of the kernel is dictated by analogy with single component diffusive aggregating systems (where the Brownian kernel is approximated by a constant kernel \cite{Chandrasekhar43,Krapivsky2011}) and by distinguishing clusters not only by their total size or mass, but by the number of particles and the number of linkers that it contains.  
 We find that, even under these strong approximations, the dynamics depends on three relevant factors, all corresponding to controlable experimental quantities: (i) the ratio between the diffusion coefficients of the particles and linkers (i.e. the existence of at least two time scales); (ii) the concentration or the relative number of particles and linkers and (iii)  the maximum number of linkers that may bond to a single particle. This enables a direct comparison between the results of the theory and existing experimental results where the effect of these control parameters was investigated \cite{Ghofraniha2009}.

The paper is organized as follows. In Sec.~\ref{sec.model}A we describe and justify
the choice of the kernel  for linker-mediated aggregation and derive the corresponding Smoluchowski equations. 
We solve these equations in two ways: by simulating a lattice model of linkers and particles whose time evolution is determined by those equations - Sec.~\ref{sec.model}B -, and analytically, using a scaling hypothesis that leads to an approximate analytical solution of the same equations - Sec.~\ref{sec.model}C. We report results for the asymptotic limit,  for the time evolution of both the bonding fraction and the size distribution of the clusters, obtained from the simulations and from the approximate analytical theory, in Sec.~\ref{sec.results}A-C. Some of the surprising experimental results reported in 
\cite{Ghofraniha2009} 
are revisited and 
%found to agree with the predictions of the approximate %analytical theory 
explained in Sec.~\ref{sec.results}D.
Finally, we discuss our results, draw some conclusions and elaborate on the possibility of generalizing the theory  in Sec.~\ref{sec.conclusion}.

\section{Theory} \label{sec.model}
\subsection{Smoluchowski equations}

We consider a mixture of $N_P$ particles and of $N_L$ linkers in a volume $V$. Particles, linkers  and clusters diffuse and bond irreversibly upon encounter. 
The linkers mediate the bonding of the particles: two particles become bonded when one of them bonds to a linker that is already bonded to another particle. Two bonded particles belong to the same cluster (and this defines a cluster as the set of particles that belong to it). Each linker bonds to a maximum of two particles and each particle bonds to a maximum of $f$ linkers ($f$ is often called the particle valence). 
As a consequence, each linker can be in one of three states: not bonded to any particle (state $0$, free linker), bonded to one particle only (state $1$) or bonded to two particles (state $2$). $N_i$ is the number of linkers in state $i$ (so that $N_0+N_1+N_2=N_L$). 
%{\xout{In this model}} 
The linkers control the extent of aggregation of the particles, and to quantify this effect we use the parameter $\phi$, 
 the ratio between the actual number of linkers and the maximum number of linkers that could be bonded to all particles,
\begin{equation}
\label{defphi}
\phi=\frac{N_L}{fN_P}.
\end{equation}
 
% {\xout{In what follows we focus on the dependence on three factors: (i) the difference in the diffusion coefficients of particles and linkers; (ii) the %relative number of particles and linkers (i.e. the role of $\phi$ in the dynamics) and (iii) the particle valence $f$.}
  
\begin{figure*} [htb]
\includegraphics[width=1.7\columnwidth]{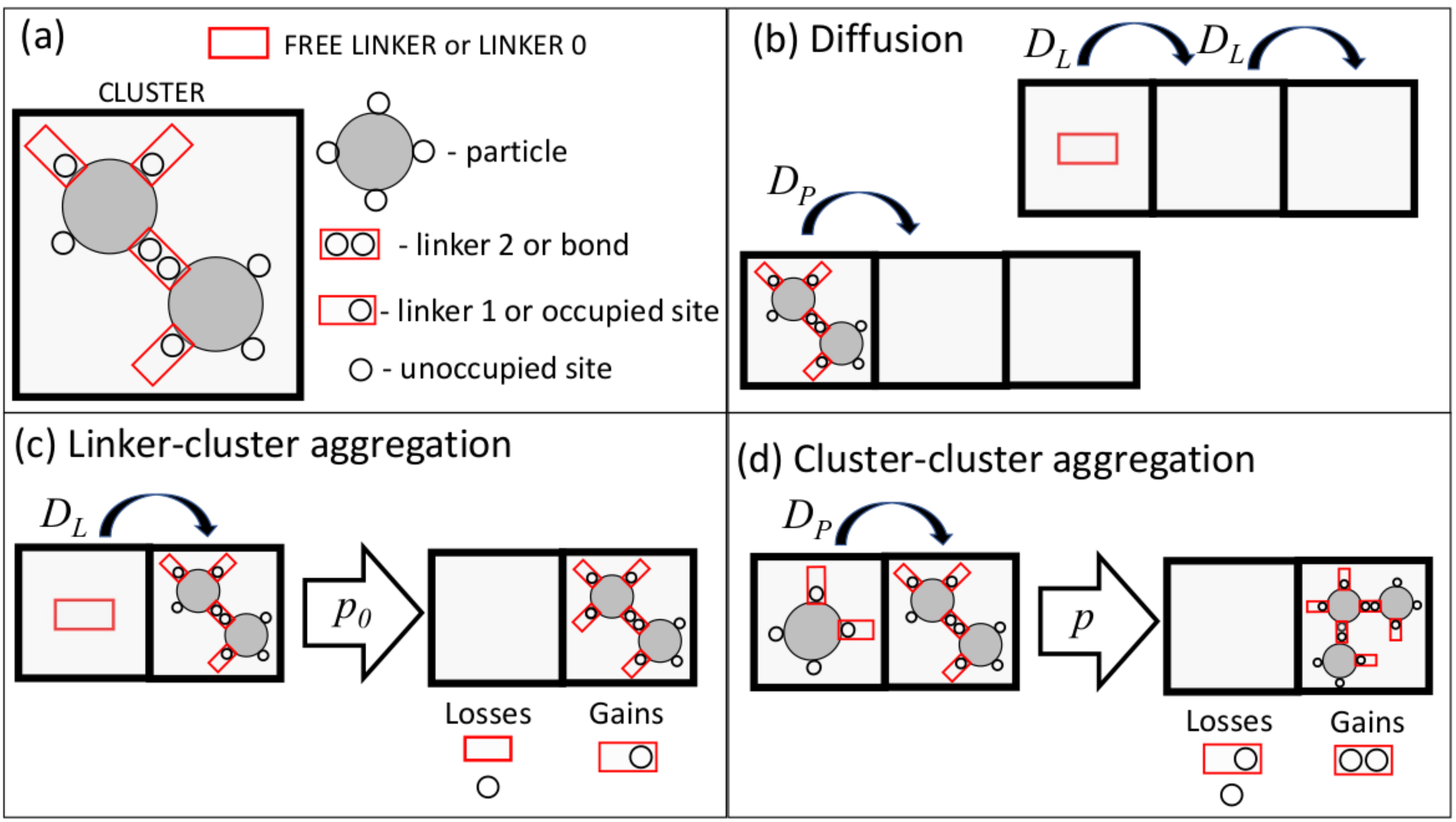}
\caption{Schematic representation of the 
%{\xout{model}} 
aggregation processes. (a) The constituents of the system are free linkers (or linkers in state 0) and clusters. A cluster $(i,j)$ consists of $i$ particles (each can be bonded to a maximum of $f$ other particles) and has $j$ sites occupied (out of $w_i=(f-2)i+2$), and 
$w_i-j$ sites unoccupied; since the cluster is tree like, it has $i-1$ bonds. An occupied site is a linker in state 1 and a bond is a linker in state 2. In the example of the figure $f=4$, $i=2$ and $j=3$; therefore $w_i=6$, there is 1 bond (or linker 2) and 3 unnocupied sites.
(b) Diffusion. The constituents diffuse randomly: linkers with a diffusion coefficient $D_L$ (or hopping rate $H_L$ in the simulation) and clusters (of all sizes) with a diffusion coefficient $D_P$ (or hopping rate $H_P$ in the simulation). Only cases where $D_L\ge D_P$ are considered. (c) Linker-cluster aggregation happens with probability $p_0$ (given by Eq. (\ref{p0ij})) when a free linker encounters a cluster $(i,j)$. After bonding to the cluster, the free linker turns into a linker 1 while the cluster gains an occupied and looses an unnoccupied site. The same bonding happens with equal probability when a cluster encounters a free linker. (d) Cluster-cluster aggregation happens with probability $p$ (given by Eq. (\ref{pi1j1i2j2})) when two clusters meet. The size of the new cluster is the sum of the sizes of the clusters that merged; as a consequence one occupied and one unoccupied sites (one in each cluster) disappear and a new bond forms. In the example of the figure, clusters $(1,2)$ and $(2,3)$  aggregate to form a cluster $(3,4)$.
 ~\label{fig.model}}
\end{figure*}

The size $i$  of a cluster is defined as the number of particles that belong to it.  We assume that two particles that are bonded to each other are connected by only one linker (no double bonding) and that the clusters are tree like (no loops). Therefore, a cluster of size $i$   contains $i-1$ linkers in state $2$, corresponding to the number of bonds needed to form it. Each cluster of size $i$
contains also a number $j$ of linkers in state 1 in the interval $[0;w_i]$, with $w_i=(f-2)i+2$. 
Notice that a particle with no linkers bonded to it is a cluster of size $i=1$ with $j=0$. 
%For the detailed description of the aggregation process
It is useful to consider that a cluster of size $i$ has a total of $w_i$ bonding sites on its ``surface''; if it contains $j$ linkers in state 1, then $j$ of these sites are occupied (by a linker) and $w_i-j$ sites are unoccupied. Therefore each cluster is characterized by two integer numbers and thus a cluster $(i,j)$ consists of $i$ connected particles and $j$ linkers in state 1 or $j$ occupied sites (see Fig. \ref{fig.model}(a)). Two clusters will aggregate when one unoccupied site in one cluster bonds to an occupied site in the other.

We now propose a generalized Smoluchowski equation for the time evolution of the clusters, which takes into account two distinct aggregation mechanisms:  bonding of free linkers to clusters $(i,j)$ and cluster-cluster aggregation.
Let $m_{ij}$ be the number of clusters $(i,j)$.
%size $i$ (i.e. formed by $i$ particles) that have $j$ linkers in state 1 bonded to the monomers ($j$ takes values from $0$ to $w_i=(f-2)i+2$). 
The rate at which $m_{ij}$ changes is,
\begin{equation}
\label{dotmijgen}
\dot m_{ij}={\cal K}_{0,+}-{\cal K}_{0,-}+{\cal K}_{+}-{\cal K}_{-},
\end{equation}
where
:
\begin{itemize}
\item[1.]{
${\cal K}_{0,+}$ is the rate at which free linkers bond to clusters $(i,j-1)$ and form a  $(i,j)$ cluster,
 \begin{equation}
 \label{K0+}
{\cal K}_{0,+}=
%(\kappa_P+\kappa_L) p_0(i,j-1) 
K_0(i,j-1)\frac{N_0 m_{i(j-1)}}{V};
 %k_0 \frac{\rho}{N_P}  p_0(i,j-1) N_0 m_{i(j-1)}
 %\frac{w_i-(j-1)}{w_{i}} m_{i(j-1)};
 \end{equation}
 }
 \item[2.]{ 
 ${\cal K}_{0,-}$ is the rate at which free linkers bond to $(i,j)$ clusters, 
 \begin{equation}
 \label{K0-}
 {\cal K}_{0,-}=
 %(\kappa_P+\kappa_L) p_0(i,j) 
 K_0(i,j)\frac{N_0 m_{ij}}{V};
 %k_0 \frac{\rho}{N_P} N_0 \frac{w_i-j}{w_{i}} m_{ij};
 \end{equation}
 }
 \item[3.]{ 
 ${\cal K}_{+}$ is the rate at which two clusters bond to form a $(i,j)$ cluster,
 \begin{equation}
 \label{K+}
 {\cal K}_{+}=
 %\kappa_P 
 \frac{1}{2}\sum_{i_1+i_2=i}\sum_{j_1+j_2=j+1}
 %p(i_1,j_1;i_2,j_2)
 K(i_1,j_1;i_2,j_2)
 \frac{m_{i_1j_1}m_{i_2j_2}}{V};
 %k_1\frac{\rho}{N_P}\sum_{i_1+i_2=i}\sum_{j_1+j_2=j+1} 
 %\frac{j_1(w_{i_2}-j_2)+j_2(w_{i_1}-j_1)}{w_{i_1}w_{i_2}}m_{i_1j_1}m_{i_2j_2};
 \end{equation}
 }
 \item[4.]{${\cal K}_{-}$ is the rate at which a  cluster $(i,j)$ bonds to other clusters:
\begin{equation}
\label{K-}
 {\cal K_{-}}=
 %2\kappa_P
 \sum_{i_1=1}^\infty\sum_{j_1=0}^{w_{i_1}}
% p(i_1,j_1;i,j)
 K(i_1,j_1;i,j)
 \frac{m_{i_1j_1}m_{ij}}
 {V}.
 %k_1\frac{\rho}{N_P}\sum_{i_1=1}^\infty\sum_{j_1=0}^{w_{i_1}}
 %\frac{j_1(w-j)+j(w_{i_1}-j_1)}{w_{i_1}w}m_{i_1j_1}m_{ij}.
  \end{equation}
  }
  \end{itemize}
$K_0(i,j)$ and $K(i_1,j_1;i_2,j_2)$ are the kernels for the two types of aggregation. They are chosen to be the product of the Brownian kernel
\cite{Chandrasekhar43,Krapivsky2011} (since we are assuming that particles, linkers and clusters diffuse between encounters)  and of a term that accounts for the fact that bond formation in an encounter depends on the number of occupied and unnoccupied sites. The Brownian kernel is approximated by a constant (see \cite{SM}), and as a consequence
%\begin{equation}
%\label{kernel1}
%K(i_1,j_2;i_2,j_2)=\left(D_{i_1j_1}+D_{i_2j_2}\right) \frac{R_{i_1j_1}+R_{i_2j_2}}{2}\times p(i_1,j_1;i_2,j_2),
%\end{equation}
%and
%\begin{equation}
%\label{kernel0}
%K_0(i,j)=\left(D_{ij}+D_{L}\right) \frac{R_{ij}+R_{L}}{2}\times p_0(i,j),
%\end{equation}
%where $D_{ij}$, $R_{ij}$ are the diffusion coefficient and the size of a cluster $(i,j)$, respectively; $D_L$, $R_L$ 
%are the diffusion coefficient and the size of linkers, respectively; and $p_(i_1,j_2;i_2,j_2)$ and $p_0(i,j)$ are probabilities of bond formation (see %below).
%The diffusion coefficients and sizes of the clusters are approximated like in \cite{Chandrasekhar43,Krapivsky2011} to obtain a constant-like kernel:
%$D_{ij}=D_P$ and $R_{ij}=R_P$, i.e., all clusters have the same diffusion coefficient and the same size of one single particle. Within this approximation the kernels become,
\begin{equation}
\label{kernel1ap}
K(i_1,j_2;i_2,j_2)=2 \kappa_P \times p(i_1,j_1;i_2,j_2),
\end{equation}
and
\begin{equation}
\label{kernel0ap}
K_0(i,j)=2\kappa_P \alpha   \times p_0(i,j),
\end{equation}
where: $\kappa_P=4\pi D_PR_P$; $\alpha=\left(1+\frac{D_L}{D_P}\right) \left(1+\frac{R_L}{R_P}\right)/4$;
 $D_L$, $R_L$ are the diffusion coefficient and the size of linkers, respectively; and $D_P$, $R_P$ are the diffusion coefficient and the size of the particles.
$p$ and $p_0$  are bonding probabilities defined in what follows.
A free linker may form a bond with a cluster $(i,j)$ if the latter has unoccupied sites. We define the probability $p_0(i,j)$ as the fraction of unoccupied sites of the cluster,
\begin{equation}
\label{p0ij}
p_0(i,j)=\frac{w_i-j}{w_i}.
\end{equation}
A cluster $(i_1,j_1)$ and a cluster $(i_2,j_2)$
  can aggregate if a bond is formed between an unoccupied and an occupied site of each cluster. Therefore, we define the 
  probability $p(i_1,j_1;i_2,j_2)$ as the product of  the fraction of unoccupied and  occupied sites of each cluster,
\begin{equation}
\label{pi1j1i2j2}
p(i_1,j_1;i_2,j_2)=\frac{j_1(w_{i_2}-j_2)+j_2(w_{i_1}-j_1)}{w_{i_1}w_{i_2}}.
\end{equation}

The restrictions to aggregation expressed by Eqs. (\ref{p0ij}) and (\ref{pi1j1i2j2}) incorporate in the dynamics the limited number of linkers that a particle can support in a simple way.
In both cases, each aggregation event is considered to be instantaneous: the formation of a bond is much faster than all other processes and may be considered instantaneous compared to the other relevant time scales (diffusion limited aggregation). 
In addition, since we focus on the limit of low density of particles and linkers, the timescale of the linker-mediated aggregation is sufficiently large for the shape and orientation of each cluster to be uncorrelated between successive bonding events. As a consequence (as already expressed in (\ref{p0ij}) and (\ref{pi1j1i2j2})), every site of the ``surface'' of a cluster  is equally likely to form the next bond.    
The relevant time scales are set by the  diffusion coefficients of the particles, $D_P$, and the free linkers, $D_L$.
%{\xout{(see Fig. \ref{fig.model}(b)). 
%For simplicity, all clusters have the same diffusion coefficient (i.e. $D_P$).}}
We define as control parameter the ratio $\Delta$ of the diffusion coeficients,
\begin{equation}
\label{Deltadef}
\Delta=\frac{D_L}{D_P}.
\end{equation}
We consider the limit where the linkers diffuse faster than the particles (i.e. $\Delta \ge 1$), since it is reasonable to assume that the linkers have a lower mass (and size) than the particles. 

%{\xout{Finally, since we consider the low density limit, it is assumed that a free linker and a cluster (of any size) occupy the same volume, thus %setting a single and trivial length scale. Notice that, within this approximation, the density is simply an initial condition for the dynamics, as the %aggregation process will effectively decrease it. }} 

The generalized Smoluchowski equation that describes cluster formation is then given by Eq. (\ref{dotmijgen}), within the approximations expressed by Eqs. (\ref{kernel1ap}),( \ref{kernel0ap}), (\ref{p0ij})  and (\ref{pi1j1i2j2}). 
The description of the dynamics of aggregation is completed with the equation for  the time evolution of the free linkers, i.e. the rate at which free linkers bond to clusters 
%{\xout{(thus changing their state from $0$ to 1)}},
 that, subject to the approximations described  is,
\begin{equation}
\label{dotN0}
\dot N_0\equiv-\sum_{i=1}^\infty \sum_{j=0}^{w_i} K_{0,-}=-2\kappa_P\alpha N_0\sum_{i=1}^\infty\sum_{j=0}^{w_i} p_0(i,j) \frac{m_{ij}}{V}.
%-k_0 \frac{\rho}{N_P} N_0\sum_{i=1}^\infty\sum_{j=0}^{w_i}
 %\frac{w_i-j}{w_{i}} m_{ij}
 \end{equation}
The aggregation dynamics for clusters of size $i$ can be obtained from Eq. (\ref{dotmijgen}) by using $m_i=\sum_{j=0}^{w_i}m_{ij}$ and taking into account that $\sum_{j=0}^{w_i}({\cal K}_{0,+}-{\cal K}_{0,-})=0$,
\begin{equation}
\label{dotmigen}
\begin{split}
\dot m_i=& \frac{\kappa_P}{V}\sum_{i_1+i_2=i}\sum_{j_1=1}^{w_{i_1}}\sum_{j_2=1}^{w_{i_2}} p(i_1,j_1;i_2,j_2) m_{i_1j_1}
%\frac{k_1\rho}{N_P}\sum_{i_1+i_2=i}\sum_{j_1=1}^{w_{i_1}}\sum_{j_2=1}^{w_{i_2}} K(i_1,j_1;i_2,j_2) m_{i_1j_1}
m_{i_2j_2}-\\
&-\frac{2\kappa_P}{V}\sum_{i_1=1}^{\infty}\sum_{j=1}^{w_{i}}\sum_{j_1=1}^{w_{i_1}} p(i,j;i_1,j_1)m_{i_1j_1}
%\frac{k_1\rho}{N_P}\sum_{i_1=1}^{\infty}\sum_{j=1}^{w_{i}}\sum_{j_1=1}^{w_{i_1}} K(i,j;i_1,j_1)m_{i_1j_1},
m_{ij}.
\end{split}
\end{equation}

 In the next two subsections we  solve Eqs. (\ref{dotN0}) and (\ref{dotmigen}) to obtain explicitly the time evolution of the 
aggregation process in the particle linker system, within the approximations adopted. First, we will perform kinetic Monte Carlo simulations of a lattice model with the same dynamics. Then, we will use a scaling hypothesis that leads to an (almost) analytical solution, by reducing Eqs. (\ref{dotN0}) and (\ref{dotmigen}) to  a system of two  ordinary differential equations.

%Notice, however, that results can be antecipated  in some limits: if $\phi=0$ there is no aggregation since there are no linkers; for $\phi=1$ and %$\Delta\gg1$ no large cluster 

\subsection{Numerical Simulations}
$N_P$ particles (or clusters $(1,0)$) and $N_L=\phi f N_P$ free linkers are randomly distributed without overlap on a simple cubic lattice with $N_{\rm{latt}}$ sites and periodic boundary conditions, each occupying one lattice site ($R_L=R_P$ in the simulations). The initial density of particles $\rho\equiv N_P/N_{\rm{latt}}=N_P/V$ is set to a low value ($\rho=0.1$ for the simulations reported) and $f$ is chosen to be 6.
% (equal to the coordination number of the lattice). 
The diffusive motion of the clusters and free linkers is described by kinetic Monte Carlo simulations. At each iteration,
one  cluster or one free linker is chosen and attempt a hop to a randomly chosen neighbouring lattice site. Free linkers are chosen with probability $H_L N_0/Q$ and clusters of size $i$ with probability $H_P m_i/Q$, where $N_0$ and $m_i$ are the actual numbers of free linkers and of clusters of size $i$, respectively, $Q$ is the total hopping rate,
\begin{equation}
Q=H_L N_0+H_P\sum_i m_i,
\end{equation}
and $H_L$, $H_P$ are two hopping rates that can be related to the diffusion coefficients of the linkers and particles (see below).
For a linker, depending on the state (occupied or empty) of the randomly chosen neighbouring site, the hop is:
\begin{itemize}
\item[(i)] {Accepted if it is not occupied;}
\item[(ii)]{ Rejected if it is occupied by another free linker;}
\item[(iii)]{ Rejected/accepted with probability  $1-p_0(i,j)$ / $p_0(i,j)$ (given by Eq. (\ref{p0ij}))  when it is occupied by a cluster $(i,j)$; when accepted the cluster $(i,j)$ is updated to a cluster $(i,j+1)$, and the free linker is changed to a linker of type 1 contained in the new cluster.} 
\end{itemize}
Similarly, for a cluster $(i_1,j_1)$, depending on the state of the randomly chosen neighbouring site, the hop is: 
\begin{itemize}
\item[(i)]{Accepted if it is not occupied;}
\item[(ii)]{Rejected/accepted with probability $1-p_0(i,j)$ / $p_0(i,j)$ (given by Eq. (\ref{p0ij})) if it is occupied by a free linker; when accepted the cluster $(i,j)$ is updated to a cluster $(i,j+1)$  and the free linker is changed to a linker of type 1 contained in the new cluster;}
\item[(iii)]{Rejected/accepted with probability  $1-p(i_1,j_1;i_2,j_2)$ / $p(i_1,j_1;i_2,j_2)$ (given by Eq. (\ref{pi1j1i2j2})) if  it is occupied by a cluster $(i_2,j_2)$; when accepted the clusters $(i_1,j_1)$ and $(i_2,j_2)$ disappear by merging into  a cluster $(i_1+i_2,j_1+j_2-1)$.}
\end{itemize}
After each hop that results in aggregation, the new cluster occupies the site 
to which the cluster/linker hops.
Therefore each cluster/linker/particle occupies one lattice site only, and the number of occupied lattice sites decreases along the simulation.
%{\xout{Notice that, since each cluster and each free linker occupies one lattice site, the number of occupied lattice sites decreases 
%by one whenever a free linker bonds to a particle site or a linker-mediated bond is formed.}} 

The discrete hopping process is transformed into continuous diffusion by: 
(i) using the relations \cite{Krapivsky2011} 
$H_L=6D_L/a^2$ and $H_P=6D_P/a^2$, where $a$ is the lattice constant; (ii) incremeting time, at each iteration,
  by a random number that follows a Poisson distribution with average value $1/Q$. 
The values of $N_0$, $m_i$ and $Q$ are updated as a result of the agreggation that follows from the encounter of the diffusive linkers and clusters.
Notice that $H_L/H_P=D_L/D_P=\Delta$.

Simulations were performed on a lattice of size $N_{\rm{latt}}=25^3$, with 
$N_P=\rho N_{\rm{latt}}=1562$ particles; $10^3$ samples were run for each pair $(\Delta,\phi)$. For $\Delta=10$ and some values of $\phi$, runs were performed also on lattices of sizes  $N_{\rm{latt}}=16^3$ and $32^3$ at the same density $\rho$, to check for finite size effects that turned out to be negligible \cite{SM}.

\subsection{Scaling hypothesis for the cluster size distribution}
 In this section we show that it is posible to approximate the full Smoluchowski description of Eq. (\ref{dotmigen}) by an analytically solvable equation, if a scaling hypothesis for $m_{ij}$ is adopted. As demonstrated below, this scaling hypothesis is equivalent to assuming that, at any given instant, the fraction $j/\omega_i$ of occupied sites of a cluster $(i,j)$ is independent of the size $i$ of the cluster and equals the ratio between the  total number of unoccupied sites (or linkers 1) and the total number of sites of the clusters.
Inspired by  the  fact that  the kernel $p(i_1, j_1;i_2,j_2)$ in Eq. (\ref{pi1j1i2j2}) is a function of $j_1/w_{i_1}$ and $j_2/w_{i_2}$, we assume the following \textit{Ansatz} for the cluster size distribution $m_{ij}$,
\begin{equation}
\label{mijansatz}
m_{ij}(t)=m_i(t) \frac{g(j/w_i,t)}{\sum_{j=0}^{w_i}g(j/w_i,t)},
\end{equation}
where $g(x,t)$ is a time dependent scaling function. Furthermore,  the moments of this function are calculated by replacing the sum by an integral, 
\begin{equation}
\label{sumbyint}
\sum_{j=0}^{w_i}j^k g(j/w_i,t)=w_i^{k+1}\int_0^1x^k g(x,t) dx.
\end{equation}
A simple approximation for the time evolution of the cluster size distribution $m_i(t)$ is then obtained by using Eqs. (\ref{mijansatz}) and (\ref{sumbyint}) 
in Eq. (\ref{dotmigen}),
\begin{equation}
\label{dotmi}
\dot m_i(t)=\frac{2\kappa_P}{V}q(1-q)\left(\sum_{i_1+i_2=i}m_{i_1}m_{i_2}-2m_i\sum_{i_1=0}^\infty m_{i_1}\right),
\end{equation}
where,
\begin{equation}
\label{qdef}
q=\frac{\int_0^1x g(x,t) dx}{\int_0^1g(x,t) dx}.
\end{equation}
Likewise, the time evolution of the free linkers is obtained by substituting Eqs. (\ref{mijansatz}) and (\ref{sumbyint}) in Eq. (\ref{dotN0}),
\begin{equation}
\label{dotN01}
\dot N_0=-\frac{2\kappa_P}{V}\alpha N_0 (1-q)\sum_{i=1}^\infty m_{i}.
\end{equation}
Notice that the quantity $q$ has a simple physical meaning: the number of linkers in state 1 is $N_1\equiv\sum_{ij} j m_{ij}$, which may be written as, 
\begin{equation}
\label{qmeaning}
N_1=q \sum_{i=1}^\infty w_i m_i ,
\end{equation}
i.e., $q$ is the ratio of the actual number of linkers in state $1$ and its maximum value for a cluster size distribution $m_i$. 
Moreover, if Eqs. (\ref{mijansatz}) and (\ref{sumbyint}) are used to replace the numerator in Eq. (\ref{qdef}), we obtain,
\begin{equation}
\label{qi}
q=\frac{\sum_{j=0}^{w_i} j m_{ij}}{w_i m_i}\equiv\frac{N_{1,i}}{w_i m_i},
\end{equation}
which means that within the approximation the fraction of the linkers $1$ bonded to clusters of size $i$ at a given instant is independent of $i$. 
%{\color{blue} and is equal to a mean 
%{\color{blue} Therefore, the scaling hypothesis expressed by Eqs. (\ref{mijansatz}) and ({sumbyint}) is  a mean-field like assumption for the distribution of linkers 1 or occupied sites: the fraction of occupied sites on clusters of size $i$ is equal to the mean fraction of
%
It is also possible to determine the time evolution of $N_2$ using Eq. (\ref{dotmi}) (and recalling that the clusters are tree like),
\begin{equation}
\label{dotN2}
\dot N_2\equiv -\sum_{i,j} \dot m_{ij}=\frac{2\kappa_P}{V} q(1-q) (N_P-N_2)^2.
\end{equation}
The time evolution of the number of free linkers  $N_0$ can be expressed as (performing the sums in Eq. (\ref{dotN01})),
\begin{equation}
\dot N_0=-\frac{2\kappa_P}{V}(1-q)N_0(N_P-N_2).
\end{equation}
The time evolution of the fraction of linkers in state $i$, $p_i\equiv N_i/N_L$, is then given by,
\begin{numcases}{}
%\boxed{
  \label{dotp0} \dot p_0=-\alpha p_0 (1-q)(1-f\phi p_2) \\
  \label{dotp2} \dot p_2=\frac{q(1-q)(1-f\phi p_2)^2}{f\phi}\\
  \label{qp0p2} q=\phi\frac{1-p_0-p_2}{1-2\phi p_2},
%  }
\end{numcases}
%{\xout{where $\Delta=\frac{\kappa_L}{\kappa_P}=\frac{D_L}{D_P}$ is the ratio (\ref{Deltadef}) between  the two time scales and} 
where the time has been rescaled by $t \to 2\rho \kappa_P  t$. Recall that $\alpha=(1+\Delta)(1+R_L/R_P)/4$. 
These equations will be solved numerically for fixed values of $\Delta$ and $\phi$ and initial conditions $p_0(0)=1$ and $p_2(0)=0$ (i.e. all linkers are free and all clusters are $(1,0)$ at $t=0$, as in the simulations). Notice that the time evolution for the fraction of linkers in state 1 can be obtained from $p_1=1-p_0-p_2$. Similarly, the ratio between the actual number of bonds between particles, $N_2$, and the maximum number of those bonds, $fN_P/2$ (that we designate by bonding probability, $p_b$), is also  obtained as a function of time, as $p_b=2\phi p_2$.  

Finally, this approximation yields an expression for the cluster size distribution $m_i$ as a function of $p_2$. 
Dividing Eq. (\ref{dotmi}) by Eq. (\ref{dotN2}) gives,
 \begin{equation}
 \label{dmidN2}
 \frac{(1-f\phi p_2)^2}{f\phi } N_P \frac{d m_i}{d p_2}=\sum_{i_1=1}^{i-1}m_{i_1}m_{i-i_1}-2m_i\sum_{i_1=1}^\infty m_{i_1},
 \end{equation}
 which, after a change of variables to $z=1/(1-f\phi p_2)-1$, becomes, 
 \begin{equation}
 \label{dcidz}
 N_P\frac{d m_i}{d z}=\sum_{i_1=1}^{i-1}m_{i_1}m_{i-i_1}-2c_i\sum_{i_1=1}^\infty m_{i_1}.
\end{equation}
This equation is formally equivalent to that obtained for the time evolution of clusters with a constant kernel \cite{Krapivsky2011}, and its solution (for the initial condition  $m_i(z=0)=N_P\delta_{i,1}$) is,
\begin{equation}
\label{ciz}
m_i = N_P\frac{z^{i-1}}{(1+z)^{i+1}}.
\end{equation}
Using the relation between $z$ and $p_2$, the cluster size distribution $m_i$ can be expressed as a function of $p_2$,
\begin{equation}
\label{finalmi}
m_i=N_P(1-f\phi p_2)^2(f\phi p_2)^{i-1}.
\end{equation}
Since $z=0$ is equivalent to the initial condition used in the simulations ($p_2=0,p_0=1$), then Eq. (\ref{finalmi}) is the theoretical prediction for the time evolution of the cluster size distribution. Notice that, as in the aggregation with constant or polymerization kernels, the cluster size distribution depends on time through the bonding proabability only.
The total number of clusters is $M_0\equiv\sum_{i=1}^\infty m_i=N_P(1-f\phi p_2)$ and the second moment of the cluster size distribution is
$M_2\equiv\sum_{i=1}^\infty i^2m_i=N_P\frac{1+f\phi p_2}{1-f\phi p_2}$.

\section{Results}\label{sec.results}
The dynamics of the model is obtained through simulations at several values of $(\Delta,\phi)$ for $f=6$ using the same initial conditions (free linkers at $t=0$: $p_0(0)=1, p_2(0)=0$). The results are compared with theoretical calculations based on the approximations discussed in the previous sections. 
 This comparison tests the validity of those approximations. Notice that in the simulations $R_L=R_P$; so, in the theoretical calculations $\alpha=(1+\Delta)/2$ in Eq. (\ref{dotp0}). In the following we present and discuss the asymptotic regimes of the
 dynamics, the time evolution of the bonding probabilities, and the time evolution of the cluster size distribution. 

\subsection{Asymptotics \label{secresass}}
\begin{figure} [thb]
\includegraphics[width=\columnwidth]{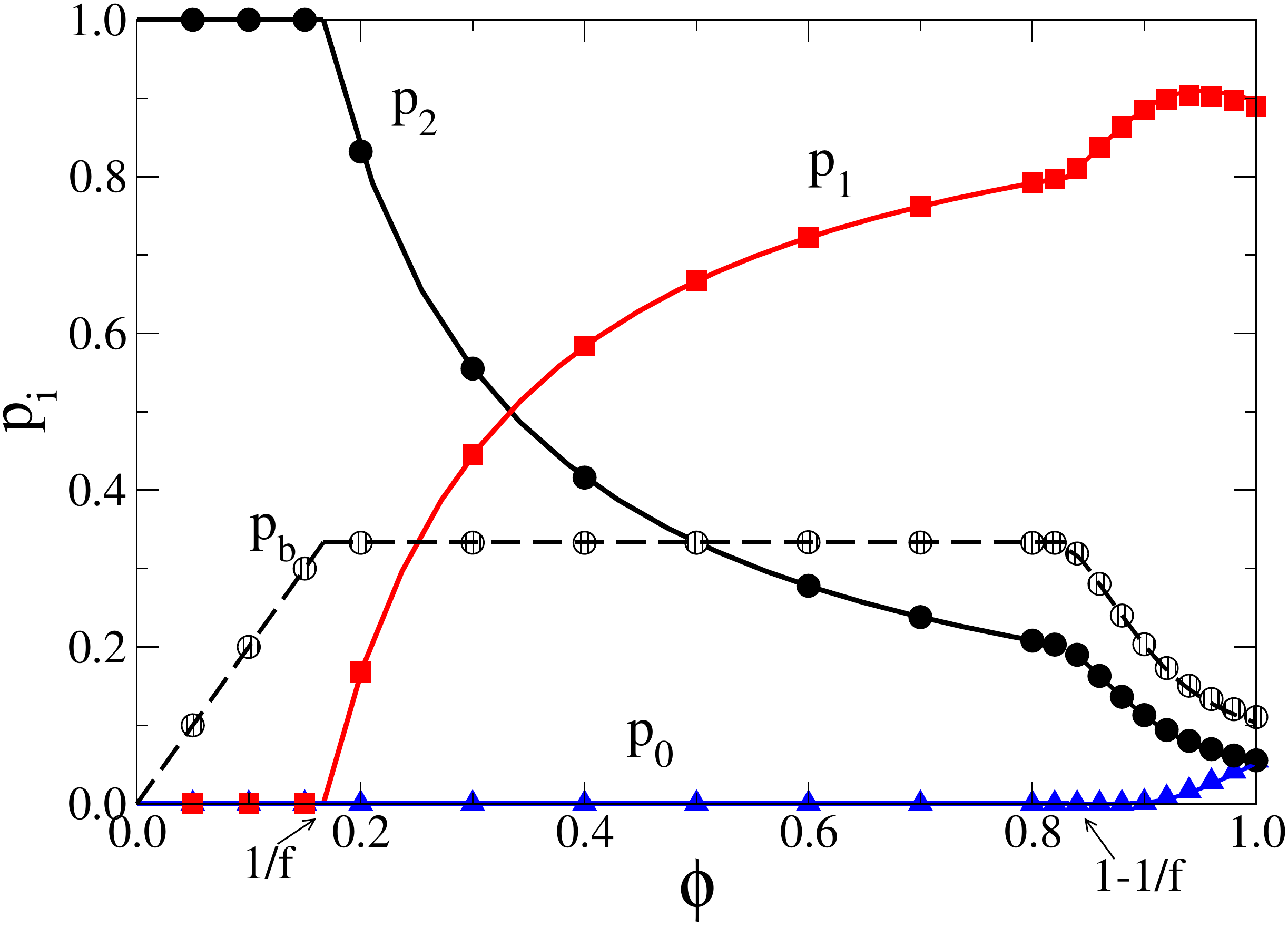}
\caption{Asymptotic values of the probabilities $p_i$ as a function of $\phi$ for $f=6$ and $\Delta=10$. The symbols represent simulation results and the lines the theoretical results from Eqs. (\ref{dotp0},\ref{dotp2}). Three different regimes are observed 
for $\phi<1/f$, $1/f<\phi<1-1/f$ and $\phi>1-1/f$. Only the latter is $\Delta$ dependent (see Fig. \ref{fig.3}).  ~\label{fig.2}}
\end{figure}	

\begin{figure} [thb]
\includegraphics[width=\columnwidth]{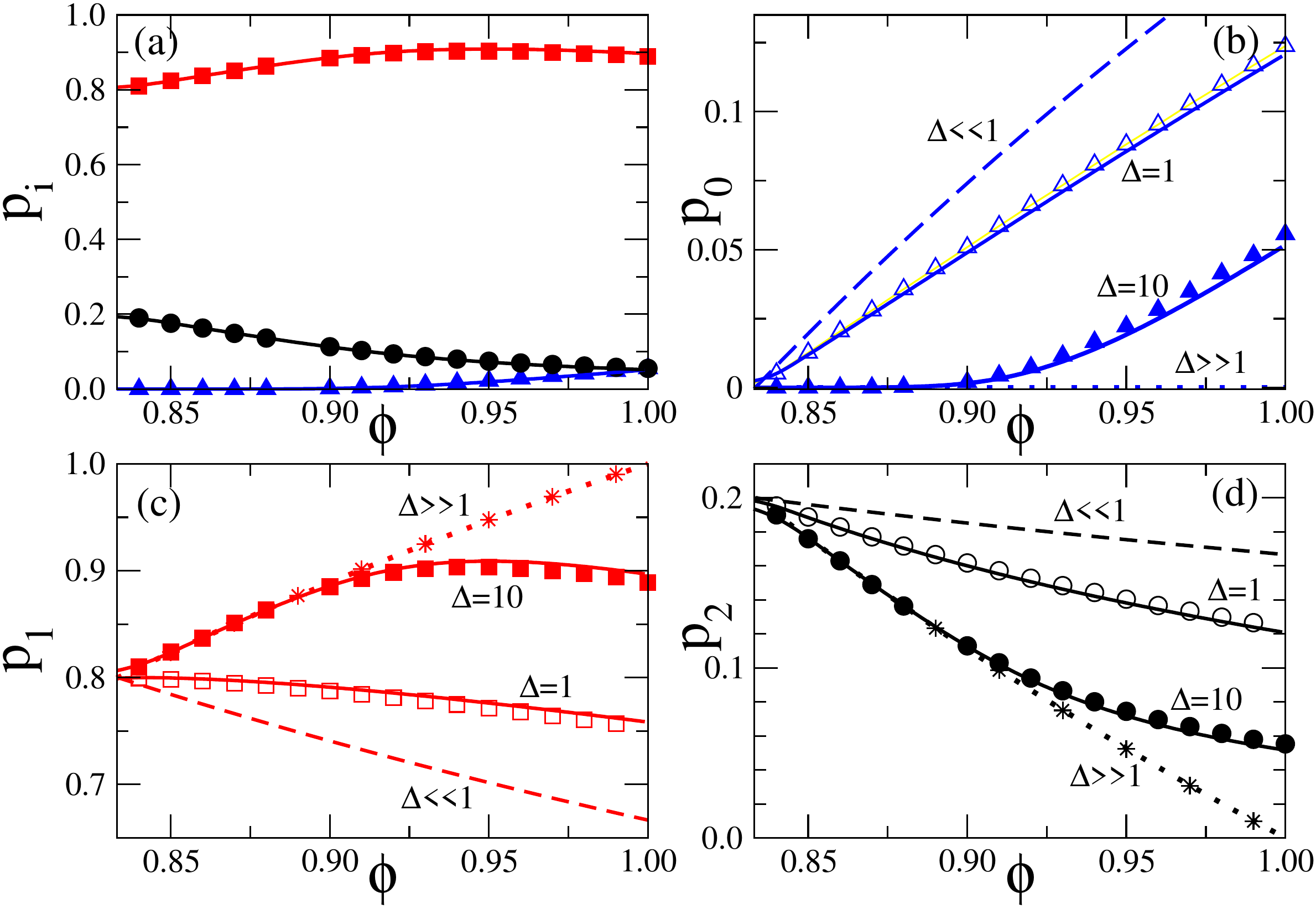}
\caption{Asymptotic values of the probabilities $p_i$ for different values of $\Delta$ in the regime $\phi>1-1/f$ (numerical results
for $f=6$). The symbols represent simulation results and the lines the theoretical results from Eqs. (\ref{dotp0},\ref{dotp2}). (a) $\Delta=10$ (a blow up of part of Fig. \ref{fig.2}); (b), (c) and (d) depict $p_0$, $p_1$ and $p_2$ respectively. The different symbols and lines correspond to different values of $\Delta$ as indicated. The analytical expressions when $\Delta \gg 1$ and $\Delta \ll 1$ may be found in the text. The asterisks in (c) and (d) are simulation results for $\Delta=100$.  ~\label{fig.3}}
\end{figure}	

The asymptotic states of the model (i.e. the values of the probabilities $p_i$ and $p_b$ when $t\to \infty$) may be found by considering the aggregation rules and the nature of the clusters. 
%We recall 
%the clusters are tree like and that the maximum number of bonds %between the particles is $N_P$ (AS IN %LINEAR CHAINS ... THIS %MAY NOT BE OBVIOUS TO A NON-EXPERT READER) 
%and thus the actual number of bonds is $N_2\leq N_P$ (or $p_2\le 1/(f\phi)$);
A tree like cluster with $i$ particles has, by definition, $i-1$ linker-mediated bonds. Thus, for the model under study, the actual number of linker-mediated bonds, $N_2$, must obey $N_2<N_P$ (or $p_2< 1/(f\phi)$). Moreover,
 since each particle can bond to at most $f$ linkers the number of occupied sites is $N_1\le f N_P$ (or $p_1\le 1/\phi$). In Fig. \ref{fig.2} we plot the asymptotic values of the bonding probabilities $p_i$ and the bonding fraction $p_b\equiv N_2/(fN_P/2)=2\phi p_2$, as a function of $\phi$ for $\Delta =10$. We find three different asymptotic regimes depending on the values of $\phi$: 

\textit{1) $0<\phi<\frac{1}{f}$.} In this regime the total number of linkers is less than the total number of particles. In fact, the number of linkers is so low that all the linkers will establish bonds between the particles (i.e. all linkers 0 become linkers 2). The time taken 
to form these bonds depends on $\Delta$  (as discussed below) but not the final state: $p_2=1$ and $p_1=p_0=0$, which is a fixed point of Eqs. (\ref{dotp0}) and (\ref{dotp2}). Due to the tree like character of the clusters, $N_2<N_P$ or $p_2\le 1/(f\phi)$, and this regime will occur when $\phi<1/f$. The asymptotic value of the bonding probability $p_b$ grows linearly with $\phi$ up to its maximum value $2/f$. These results are independent of $\Delta$ (for simplicity only $\Delta=10$ is shown in Fig. \ref{fig.2}). 

\textit{2) $\frac{1}{f}<\phi < 1-\frac{1}{f}$.} In this regime, the number of linkers is larger than the total number of particles but smaller than $N_P(f-1)$. The number of bonds between the particles takes its maximum value $p_2=1/(f\phi)$ and the bonding probability, $p_b=2/f$, is maximal and independent of $\phi$. Since $N_L> N_P$ or $\phi>1/f$, the number of type 1 linkers is non-zero $N_1=N_L-N_P$ or $p_1=1-1/(f\phi)$ (and $p_0=0$). This regime corresponds to the fixed point $p_0=0$, $p_2=1/(f\phi)$ of Eqs. (\ref{dotp0}) and (\ref{dotp2}), and will occur when $\phi< 1-1/f$, since $p_1+p_2=1\le1/(f\phi)+1/\phi$. The asymptotic state is also independent of $\Delta$, and is shown in  Fig. \ref{fig.2} for $\Delta=10$. The same results are obtained for any other value of $\Delta$. The constant bonding fraction $p_b=2/f$ implies that in this regime there is a single cluster, which contains all the particles, in the limit $t \to\infty$.

\textit{3) $\phi>1-\frac{1}{f}$.} In this regime the asymptotic state is determined by the exhaustion of unoccupied sites in the clusters. At $t=0$, the number of unnocupied sites is $fN_P$; since each linker 1 occupies one site and each linker 2 occupies two sites, the fraction of unoccupied sites at any time is  $1-\phi(p_1+2p_2)$, and vanishes when $p_1+2p_2=1/\phi$.  The values of $p_1$ and $p_2$ which satisfy this condition depend on $\Delta$. This can be seen by analysing two limiting cases. When $\Delta \gg 1$, the $N_L$ linkers, free at $t=0$, change to linkers 1 before any bonds between the clusters are formed; at this stage there are $N_L$ linkers 1 and $fN_P-N_L$ unoccupied sites that will then start to form bonds between the particles or clusters; at the end of this process, there are $fN_P-N_L$ bonds ($p_2=1/\phi-1$) and $2N_L-fN_P$ linkers 1 ($p_1=2-1/\phi$), and the aggregation stops since all the sites are occupied. 
In the other limit, when $\Delta \ll 1$, when a free linker bonds to an unoccupied site a bond between two clusters is also formed 
between the newly occupied site and an unnocupied one; this process lasts while $N_2$ remains below its maximum value $N_P$ ($p_2=1/(f\phi)$). Thereafter the remaining free linkers $N_L-N_P$ will bond with the $(f-2)N_P$ still unoccupied sites;  this second process ends when $N_1=(f-2)N_P$ (or $p_1=(1-2/f)/\phi$), as all the sites are occupied. Notice that at this stage a number of  linkers are still free as $N_0=N_L-(f-1)N_P$ (or $p_0=1-(1-1/f)/\phi$) is non-zero. The asymptotic state for finite values of $\Delta$ lies between these limits. This regime corresponds to the fixed point $q=1$ of Eqs. (\ref{dotp0}) and (\ref{dotp2}). 
In Figs. \ref{fig.2} and \ref{fig.3}(a) we show the asymptotic values of $p_i$ for $\Delta=10$ and $1-1/f<\phi<1$, obtained from the simulations and the theory, which are in quantitative agreement. Figures \ref{fig.3}(b)-(d) illustrate this for other values of $\Delta$ and reveal the same excellent agreement between the simulation and the theoretical results. 

We conclude that the asymptotic numbers of bonds and clusters depend on the ratio of the diffusion coefficients $\Delta$ only 
when $N_L>(f-1)N_P$. Faster free linkers  (larger  $\Delta$) promote the depletion of unoccupied sites decreasing the formation of bonds between the clusters. For intermediate values of $N_L$, i.e. $1/f<\phi<1-1/f$, the model predicts the formation of a single cluster, containing all the particles, in the long time limit.

 \begin{figure} [thb]
\includegraphics[width=\columnwidth]{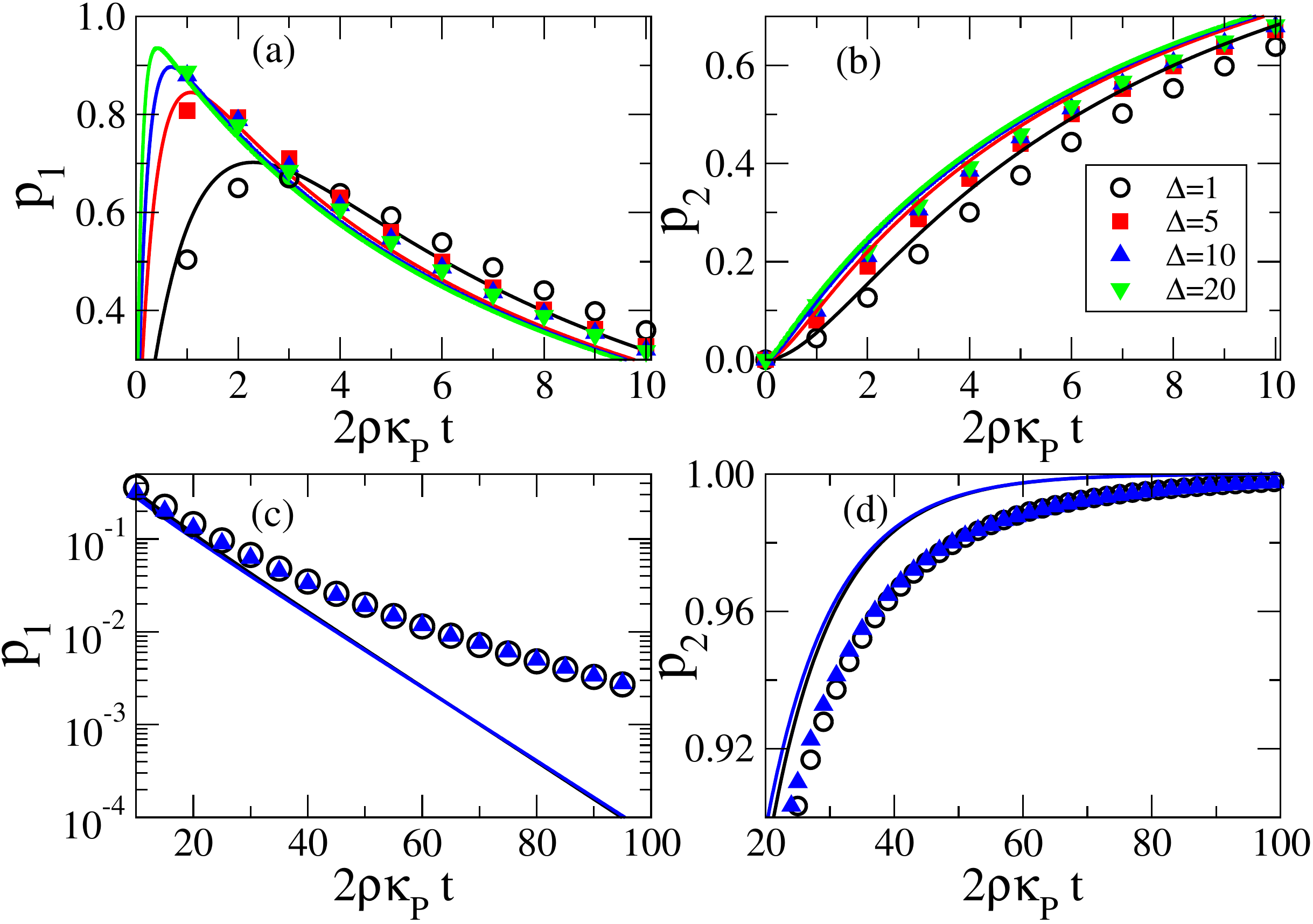}
\caption{Time evolution of the probabilities $p_1$ ((a) and (c)) and $p_2$ ((b) and (d)) at fixed $\phi=0.05$ and the indicated values of $\Delta$, at short ((a) and (b)) and long ((c) and (d)) times. The symbols represent simulation results and the lines are numerical solutions of Eqs. (\ref{dotp0},(\ref{dotp2}). For the sake of clarity, in (c) and (d) only the results for $\Delta=1$ and $\Delta=10$ are shown (the results are independent of $\Delta$). ~\label{fig.4}}
\end{figure}	
 \begin{figure} [htb]
\includegraphics[width=\columnwidth]{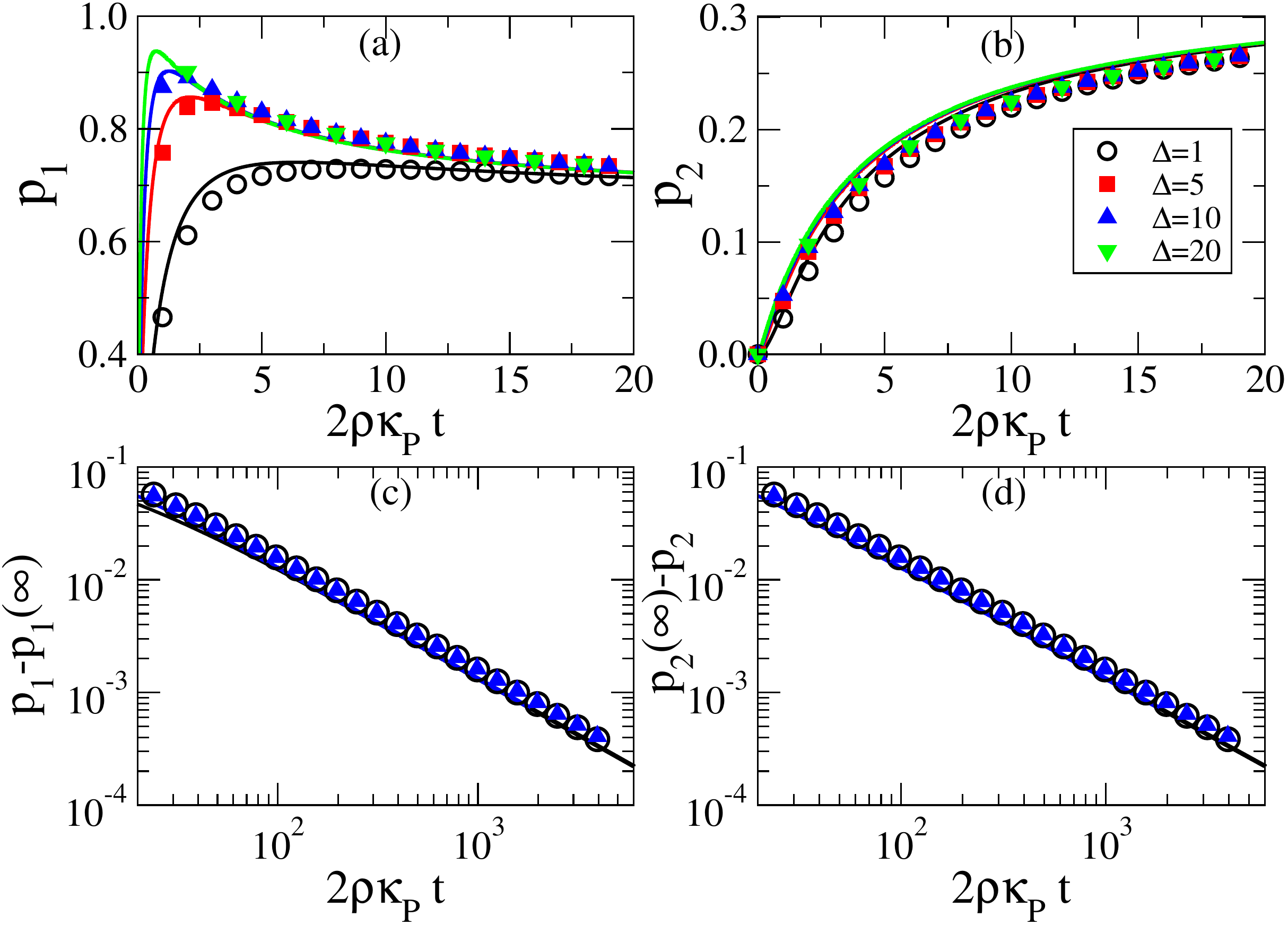}
\caption{Time evolution of $p_1$ and $p_2$ as in Fig. \ref{fig.4} at $\phi=0.5$.  ~\label{fig.5}}
\end{figure}	 
 \begin{figure} [htb]
\includegraphics[width=\columnwidth]{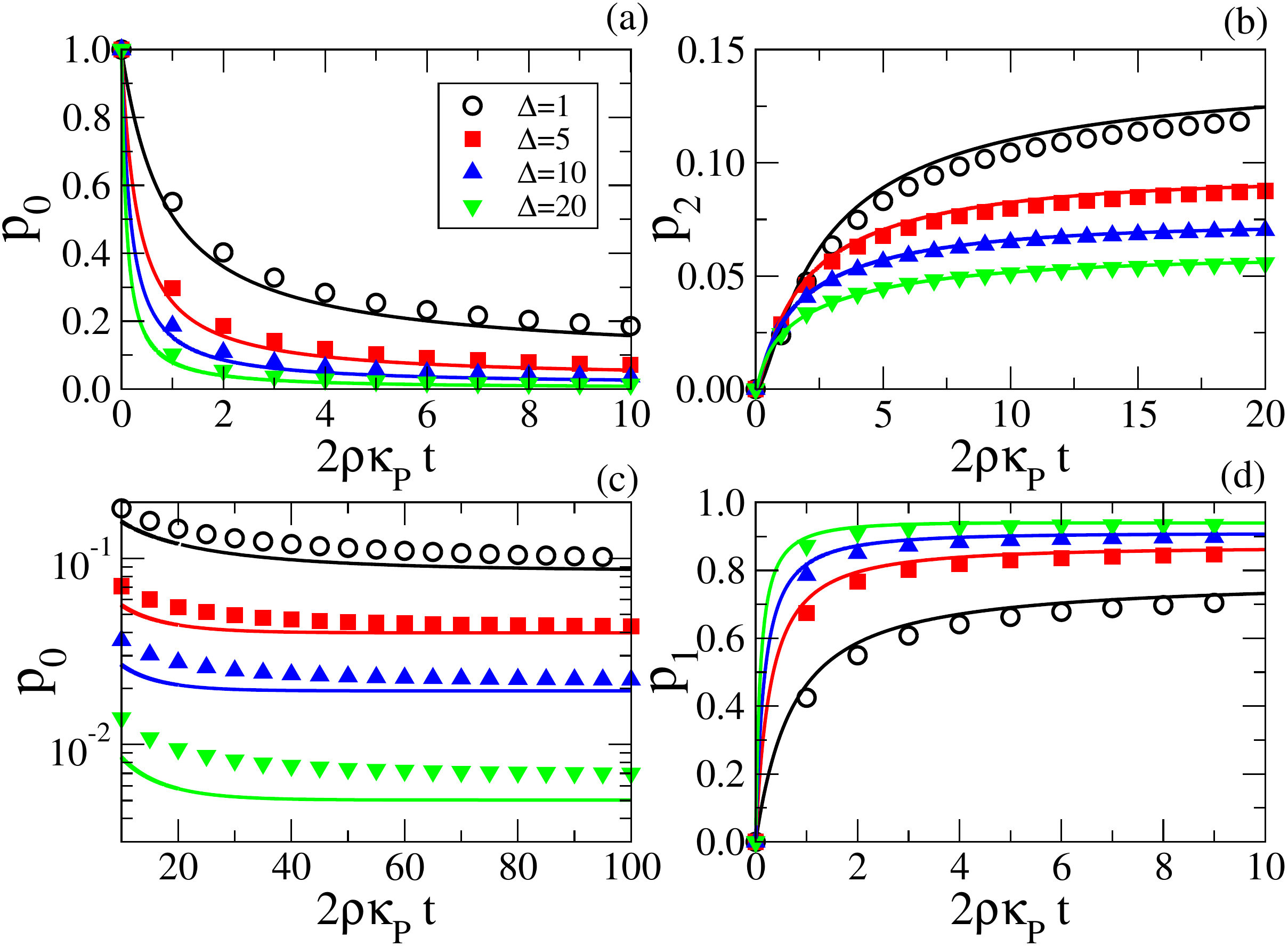}
\caption{ Time evolution of $p_0$, $p_1$ and $p_2$ as in Figs. \ref{fig.4} and \ref{fig.5} at $\phi=0.95$. ~\label{fig.6}}
\end{figure}	
\subsection{Time evolution of the bonding probabilities}
We consider the dynamics of the aggregation at three different values of $\phi=0.05,\,0.5$ and $0.95$ (representative of the three asymptotic regimes), and different diffusion ratios, $\Delta=1, \,5,\,10$ and 20. We keep $f=6$ fixed.
At the two lowest values of $\phi$, the fraction of linkers of type 1, $p_1$, exhibits a maximum at early times. The value of this maximum increases and the time when it occurs decreases as $\Delta$ increases (see Figs. \ref{fig.4}(a) and \ref{fig.5}(a)). The presence of the maximum signals the change from an initial regime dominated by the formation of bonds between free linkers and unoccupied sites to a second regime where bonds between the particles form and the clusters merge. This transition is absent at $\phi=0.95$  (see Fig. \ref{fig.6} (d)): if the fraction of free linkers is sufficiently high bond formation between free linkers and unoccupied sites dominates at all times, due to the large decrease of unoccupied sites required for the formation of bonds between clusters.
Figs. \ref{fig.4} and \ref{fig.5} show that, at low  $\phi$, the time evolution of the bonding probabilities can be divided into an initial $\Delta$ dependent regime (up to $2\rho \kappa_P t \approx 10-20$, Figs. \ref{fig.4} and \ref{fig.5} (a), (b)) and a late $\Delta$ independent regime (Figs. \ref{fig.4} and \ref{fig.5} (c), (d)). By contrast, at large values of $\phi$ the dynamics depends on $\Delta$ at all times (see Fig. \ref{fig.6}). 

The theoretical results are validated by comparing them to the results of simulations (lines and symbols in Figs \ref{fig.4}-\ref{fig.6}). At low $\phi=0.05$, initially, there is qualitative agreement for the time evolution of $p_1$ and $p_2$, including their dependence on $\Delta$ (Fig. \ref{fig.4}(a) and (b)). 
At late times, there is agreement on the prediction that the dynamics does not depend on $\Delta$ (Figs. \ref{fig.4}(c) and(d)). However, a quantitative discrepancy has been observed: while the theory predicts fast exponential decay to the asymptotic values, the simulation reveals a slower time evolution. 
This discrepancy between the simulation and the theoretical results (and also the observed in Fig. \ref{fig.6}(c)) may be attributed to fluctuations in the distribution of the fraction of occupied sites in clusters with different sizes. Indeed, the analytical theory assumes that, at each instant, the fraction of occupied sites of a cluster is independent of its size (Eq. (\ref{qi})). This means that, under conditions where most of the clusters have either a few (low $\phi$) or many (high $\phi$) occupied sites, a strongly varying quantity is approximated by its average value, giving rise to the discrepancies observed in these regimes.
At $\phi=0.5$ (Fig. \ref{fig.5}) the agreement between the theory and simulations is almost quantitative at all values of $\Delta$. Both predict a power-law decay of the evolution of the probabilities at late times, with an exponent $\approx-1$ (Figs. \ref{fig.5}(c) and (d)). This feature results from the non-linearity of Eq. (\ref{dotp2}), which for $1/f<\phi<1-1/f$, and $t \to \infty$ reduces to,
\begin{equation}
\label{dotp2ass}
\dot x= -B(f,\phi)x^2,
\end{equation}
where $x=1-f\phi p_2$ and $B(f,\phi)=(f-1-f\phi)(f\phi-1)/(f-2)^2$. The solution of this equation yields $p_1(t)-p_1(t=\infty)=1/(B f\phi t)$. This power law decay was observed in simulations for other values of $\phi$ in the range $1/f<\phi<1-1/f$, which strongly suggests that the present model in this range behaves as irreversible aggregation with a constant kernel \cite{Krapivsky2011}.  
 
 \begin{figure} [tbh]
\includegraphics[width=0.48\columnwidth]{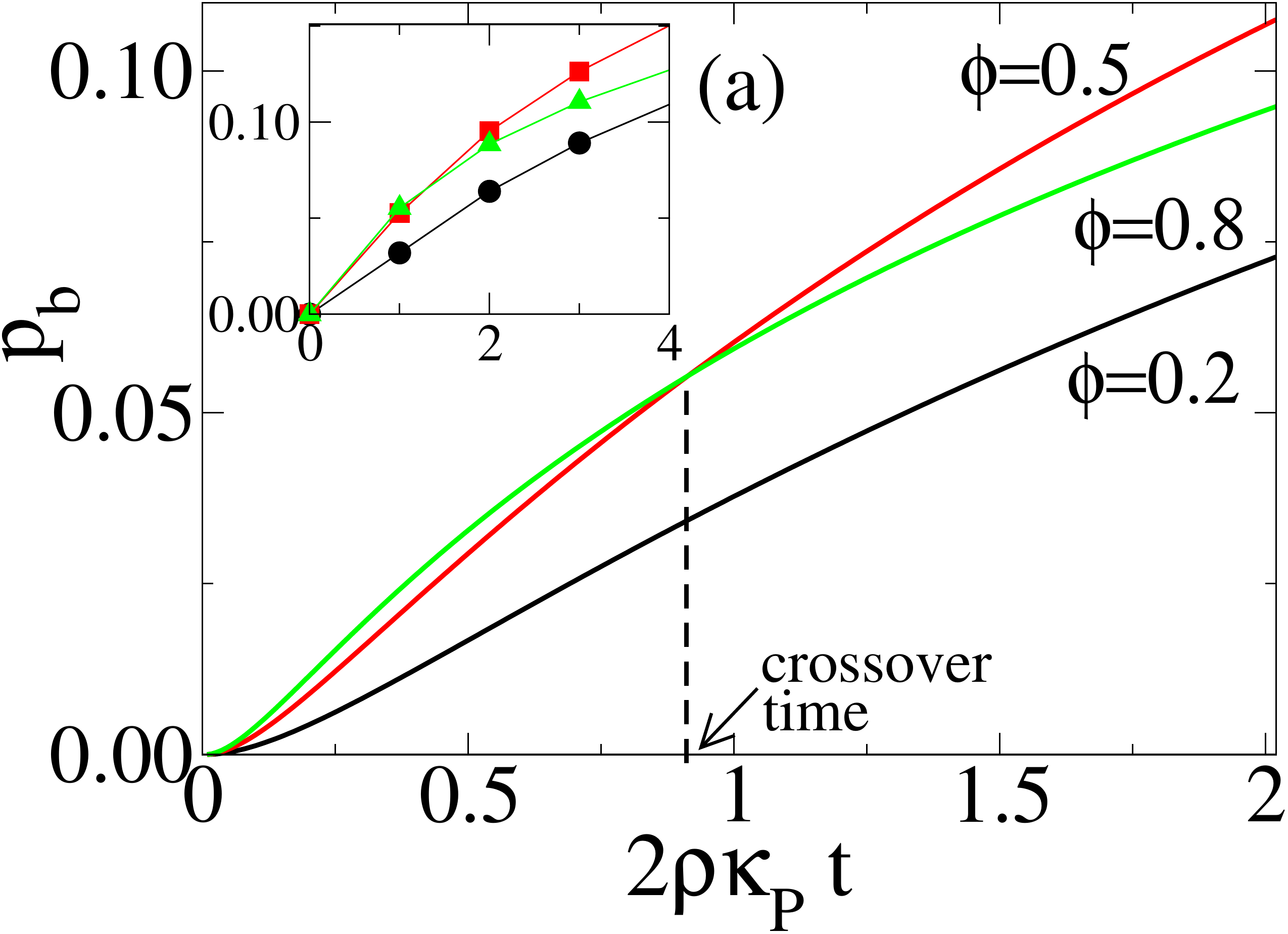}
\includegraphics[width=0.48\columnwidth]{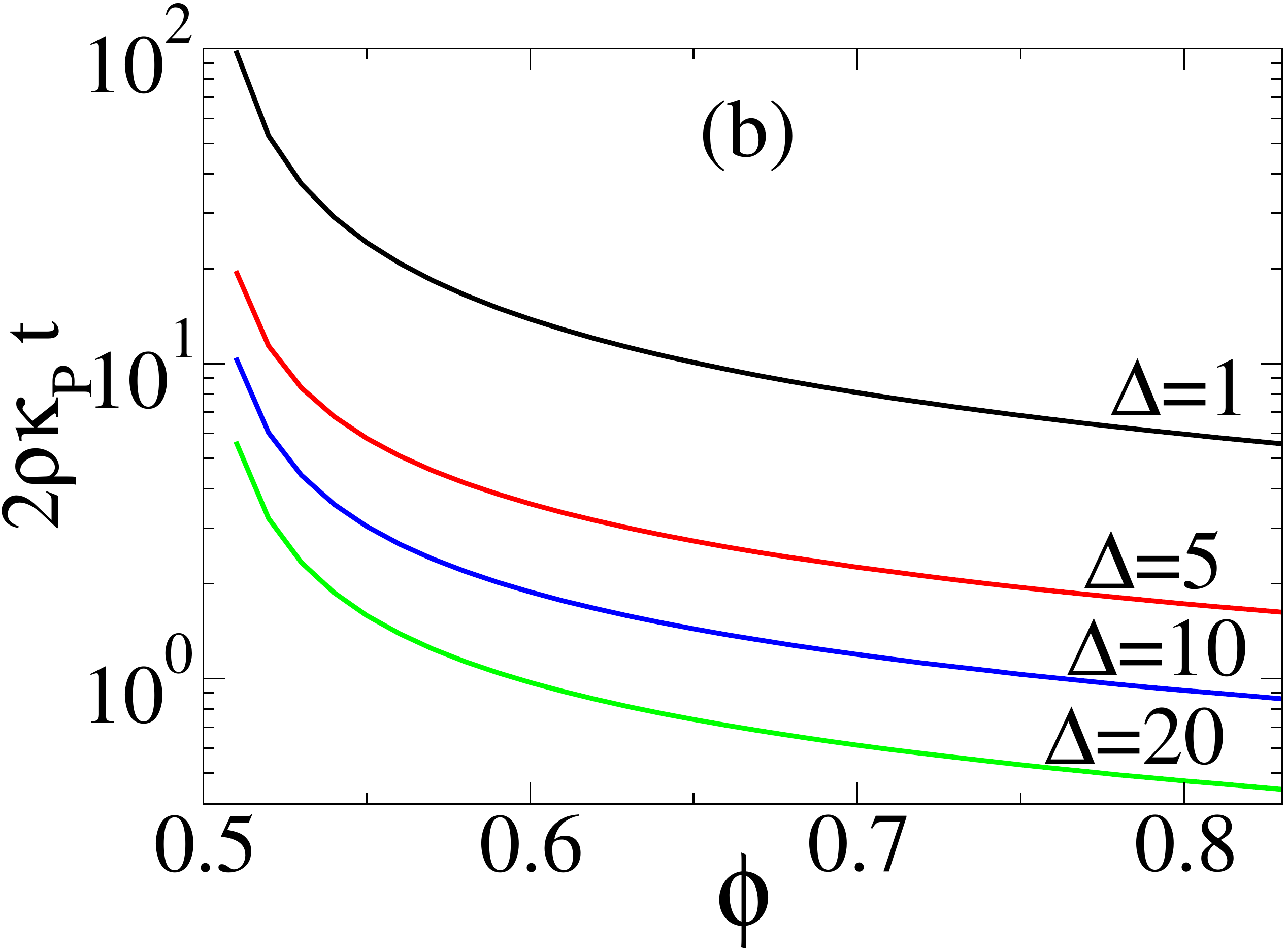}
\caption{(a)Bonding probability, $p_b$, as a function of time at the early stages for $\Delta=10$ and for three values of $\phi$. The lines represent theoretical results Eqs. (\ref{dotp0}) and (\ref{dotp2}). The crossover time, where $p_b$ at $\phi=0.5$ equals $p_b$ at $\phi=0.8$, is indicated. The inset displays the simulation results for the same parameters, namely, $\phi=0.2$ (circles), $\phi=0.5$ (squares) and $\phi=0.8$ (triangles); in the inset the lines are guides to the eye. (b) Crossover time  as a function of $\phi$ for different values of $\Delta$.~\label{fig.7}}
\end{figure}	

The maximum number of bonds between particles (mediated by type 2 linkers) is reached asymptotically whenever $1/f<\phi<1-1/f$
for any $\Delta$ as discussed in section \ref{secresass} (see Fig. \ref{fig.2}). In this regime, the system tends to form a single cluster that contains all particles. 
In order to investigate the conditions $(\phi,\Delta)$ under which the time required for the aggregation of this cluster is minimal, we analyse Eqs. (\ref{dotp0}) and (\ref{dotp2}) in the limits $t \to 0^+$ and $t\to\infty$, to obtain the behaviour of the bonding fraction $p_b$ in these limits.
At $t=0$, one has $\dot p_b =0$ and $\ddot p_b=\phi(1+\Delta)/f$, and therefore, 
\begin{equation}
\lim_{t\to0^+} p_b(t)=\phi\frac{1+\Delta}{f} t^2.
\end{equation}
As a consequence, the initial growth of the clusters increases with $\Delta$ and $\phi$. 
The limit $t \to \infty$ may be obtained using Eq. (\ref{dotp2ass}) as  $p_b=2/f(1-x)$ and thus, 
\begin{equation}
\lim_{t\to\infty} p_b(t)=\frac{2}{f}\left(1-\frac{1}{B(f,\phi)t}\right).
\end{equation}
The function $B(f,\phi)$ has a maximum at $\phi=0.5$. Thus, for any $f$ and $\Delta$, the single cluster limit will be approached faster when $\phi=0.5$. This behaviour is illustrated in Fig. \ref{fig.7}(a), where the value of $p_b$ for $f=6$ and $\Delta=10$ is plotted for $\phi=0.2,\,0.5$ and $0.8$, as obtained from the theory and simulations. 
Initially, the fastest growth is obtained for the largest $\phi$ ($\phi=0.8$). However, the bonding fractions for $\phi=0.8$ and $\phi=0.5$ cross at $t\approx 5$ (in the units used in the figure) and the bonding fraction for $\phi=0.5$ becomes the largest at later times. 
Figure \ref{fig.7}(b) represents this crossover time as a function of $\phi \in ]0.5;1-1/f]$ for several values of $\Delta$. This result may have experimental relevance: if the goal is the growth of the largest cluster in the shortest time then, for a given number of $N_P$ particles, the optimal number of linkers is $N_L=fN_P/2$, irrespective of their difusion coefficients. 

%{\color{blue} Finally, it is also interesting to compare the present results to those of a system with only particles (no linkers) that could bond directly %to each other through their $f$ bonding sites. In such a case, the theory with a constant kernel predicts that $p_b=\frac{2}{f} \frac{t}{1+t}$ \cite{Krapivsky2011}. Figure ... compares this result to those obtained with the approximate theory for particles and linkers for several values of $\Delta$ and $\phi$. Comment...} 

\subsection{Cluster size distributions}
\begin{figure} [tbh]
\includegraphics[width=\columnwidth]{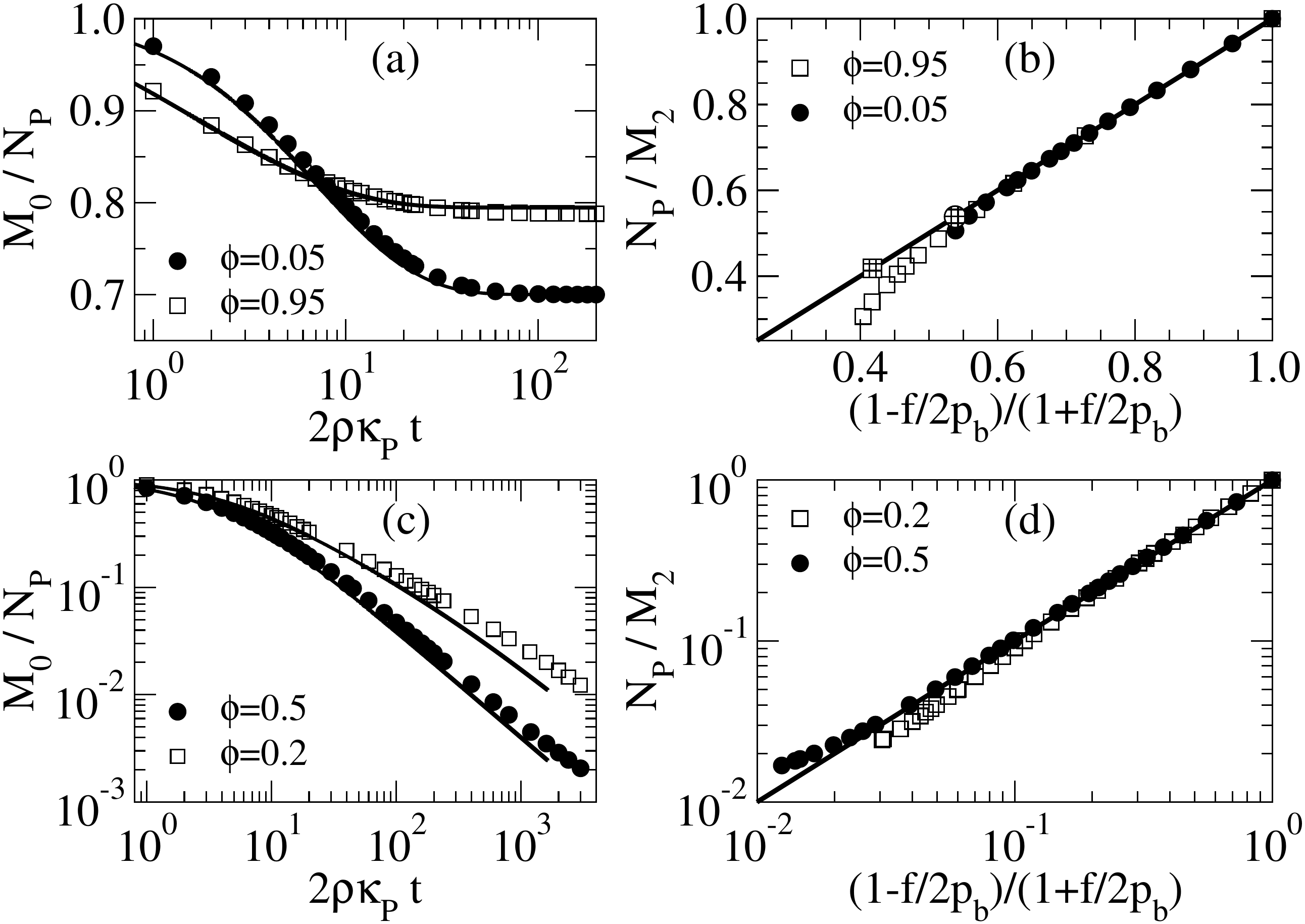}
\caption{Inverse mean cluster size, $M_0/N_P$ ((a) and (c)) as a function of time, 
and inverse second moment of the cluster size distribution ((b) and (d)), $N_P/M_2$, as a function of $(1-fp_b/2)/(1+fp_b/2)$. $\Delta=10$ and $f=6$ in all cases; the values of $\phi$ are indicated in the figures. The symbols are the results from simulations and the lines from Eqs. (\ref{dotp0}), (\ref{dotp2}) and (\ref{fracclusti}).  The  patterned symbols in (b) indicate the asymptotic values predicted by the theory at $\phi=0.95$ (square) and $\phi=0.5$ (circle). ~\label{fig.8}}
\end{figure}

In this section, we compare the results of simulations and theory for the cluster size distribution. Notice that this quantity  can be measured in more realistic simulations and even in experiments  \cite{Ghofraniha2009,Hiddessen2000}.
The total number of clusters is the zeroth moment, $M_0$, of the cluster size distribution $m_i$. It is, by definition, related  to the number of linkers in state $2$, $M_0=N_P-N_2$, since the clusters are tree like. As  a consequence, the mean cluster size, 
$\langle S_1 \rangle \equiv N_P/M_0$, can be calculated from $p_2$: $\langle S_1 \rangle=1/(1-f\phi p_2)$. In Figs. \ref{fig.8} (a) and (c), $\langle S_1\rangle^{-1}$ is represented for $\Delta=10$, $f=6$. $\langle S_1\rangle$ grows with time, but the extent of this growth depends on $\phi$. At $\phi<1/f$  it reaches the asymptotic value $\langle S \rangle =1/(1-f\phi)$ -- see Fig. \ref{fig.8}(a). %$\langle S_1\rangle^{-1}=0.7$ for $\phi=0.05$.%
At $\phi>1-1/f$, the asymptotic value of $\langle S_1 \rangle$  depends on $\Delta$. In both cases, the growth of large clusters is strongly limited as we observed $\langle S_1 \rangle <2$ at all times.
By contrast, when $1/f<\phi<1-1/f$, there is no finite asymptotic limit for $\langle S_1 \rangle$, as shown in Fig. \ref{fig.8}(c). The mean cluster size grows monotonically with time; the dependence on $\phi$ is limited to the time taken to reach a given cluster size, which has a minimum at $\phi=0.5$ (at times larger than the crossover described in Fig.~\ref{fig.7}). The results in Fig. \ref{fig.8}(a) at $\phi=0.05$ and $0.95$  reveal that the agreement between simulations and theory is good while those in Fig. \ref{fig.8}(c) at $\phi=0.5$ and $0.2$ show that the theory slightly overestimates $\langle S_1 \rangle$ at long times. 
 
The cluster size distribution was determined using simulations for $\Delta=10,\, f=6$ at several values of $\phi$ and different times.
The inverse of $M_2$, the second moment of these distributions, divided by $N_P$, is represented in Figs. \ref{fig.8}(b) and (d), as a function of $(1-fp_b/2)/(1+fp_b/2)$ to facilitate the comparison with the theoretical prediction. A behaviour similar to that of $\langle S_1 \rangle$ is observed: limited growth at $\phi=0.05$ and $0.95$ and an asymtptotic divergence at $\phi=0.2$ and $0.5$.  The theory overestimates $M_2$ in all cases except at $\phi=0.5$. 
 
\begin{figure} [tbh]
\includegraphics[width=0.45\columnwidth]{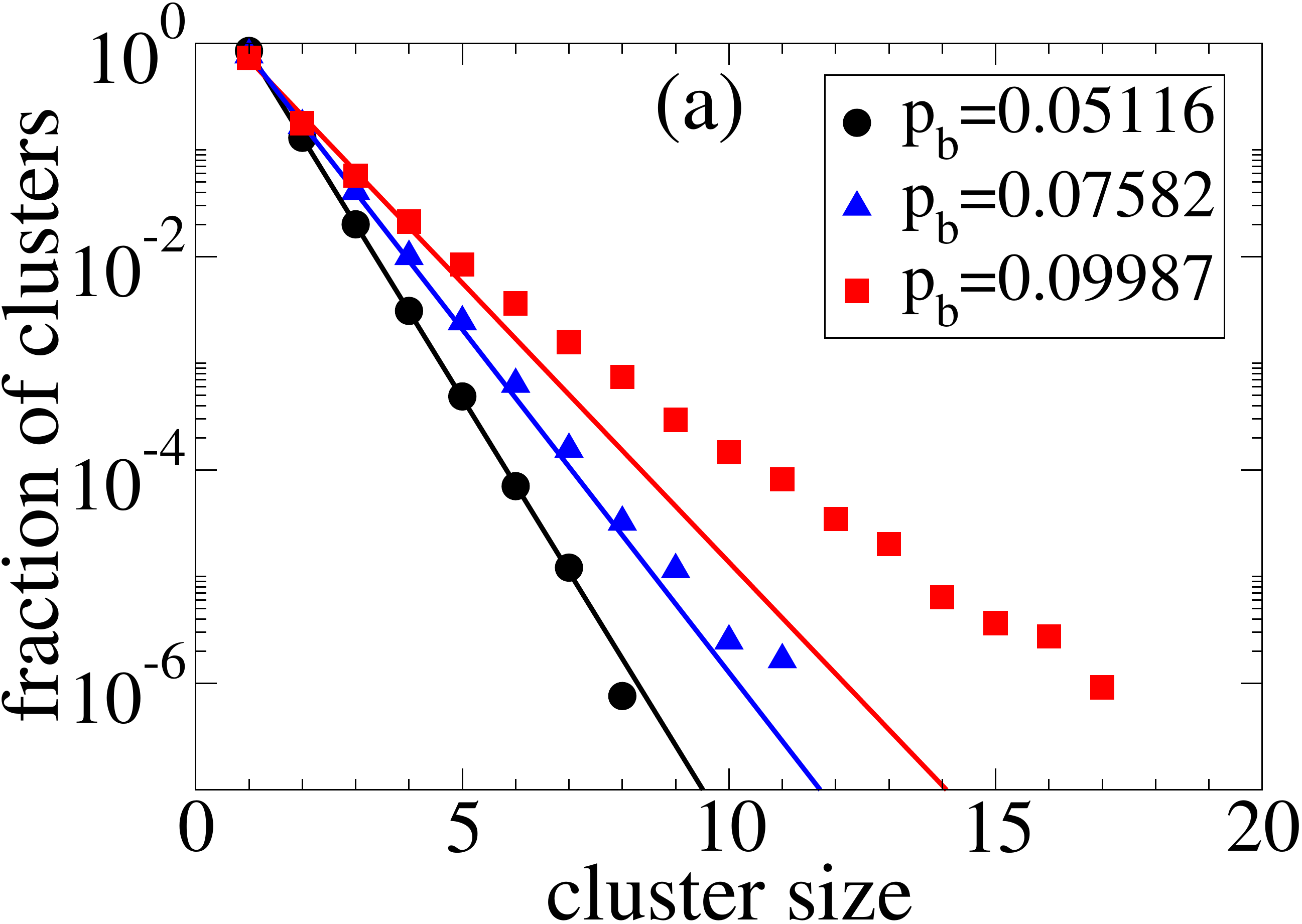}
\includegraphics[width=0.45\columnwidth]{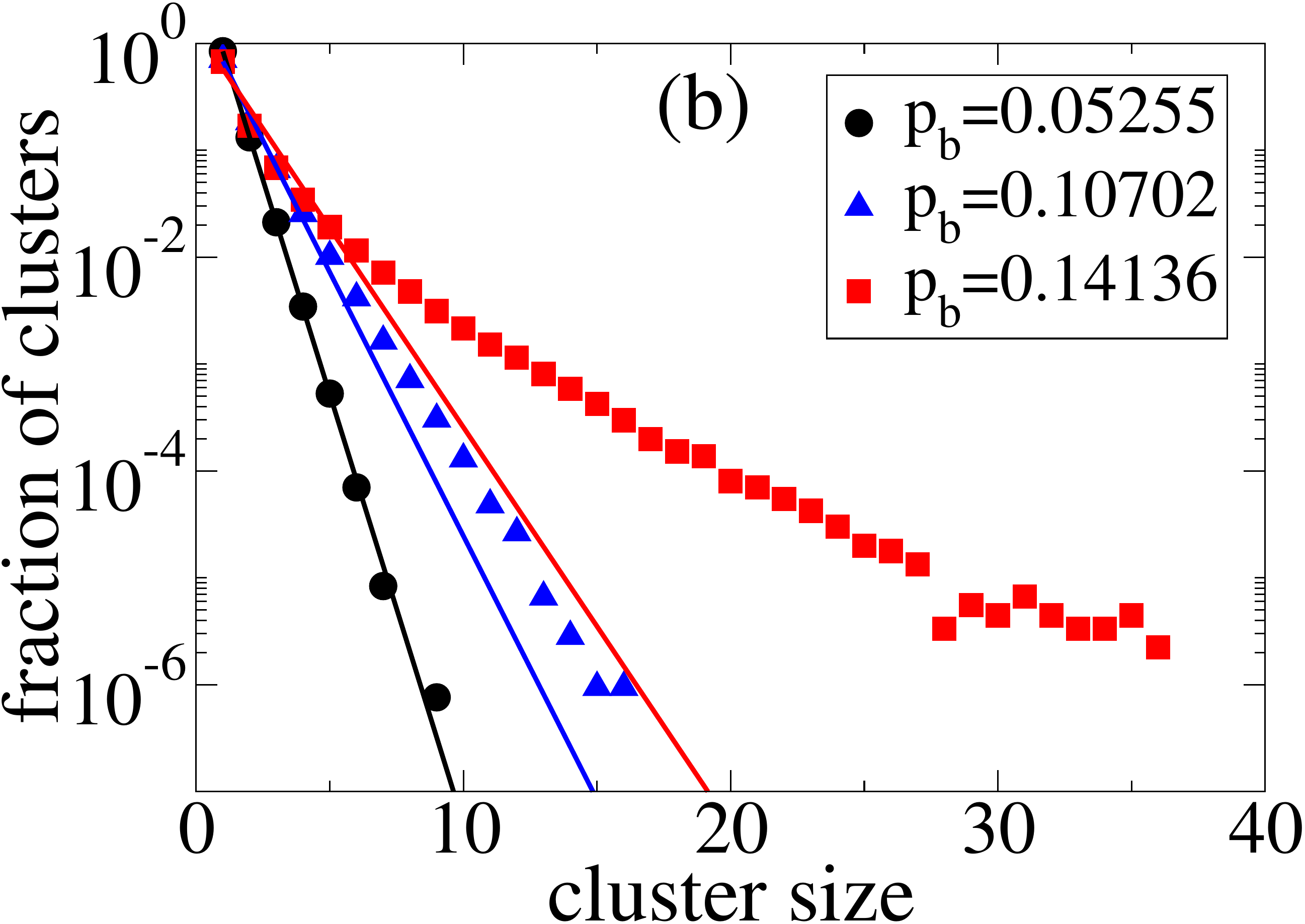}\\
\includegraphics[width=0.45\columnwidth]{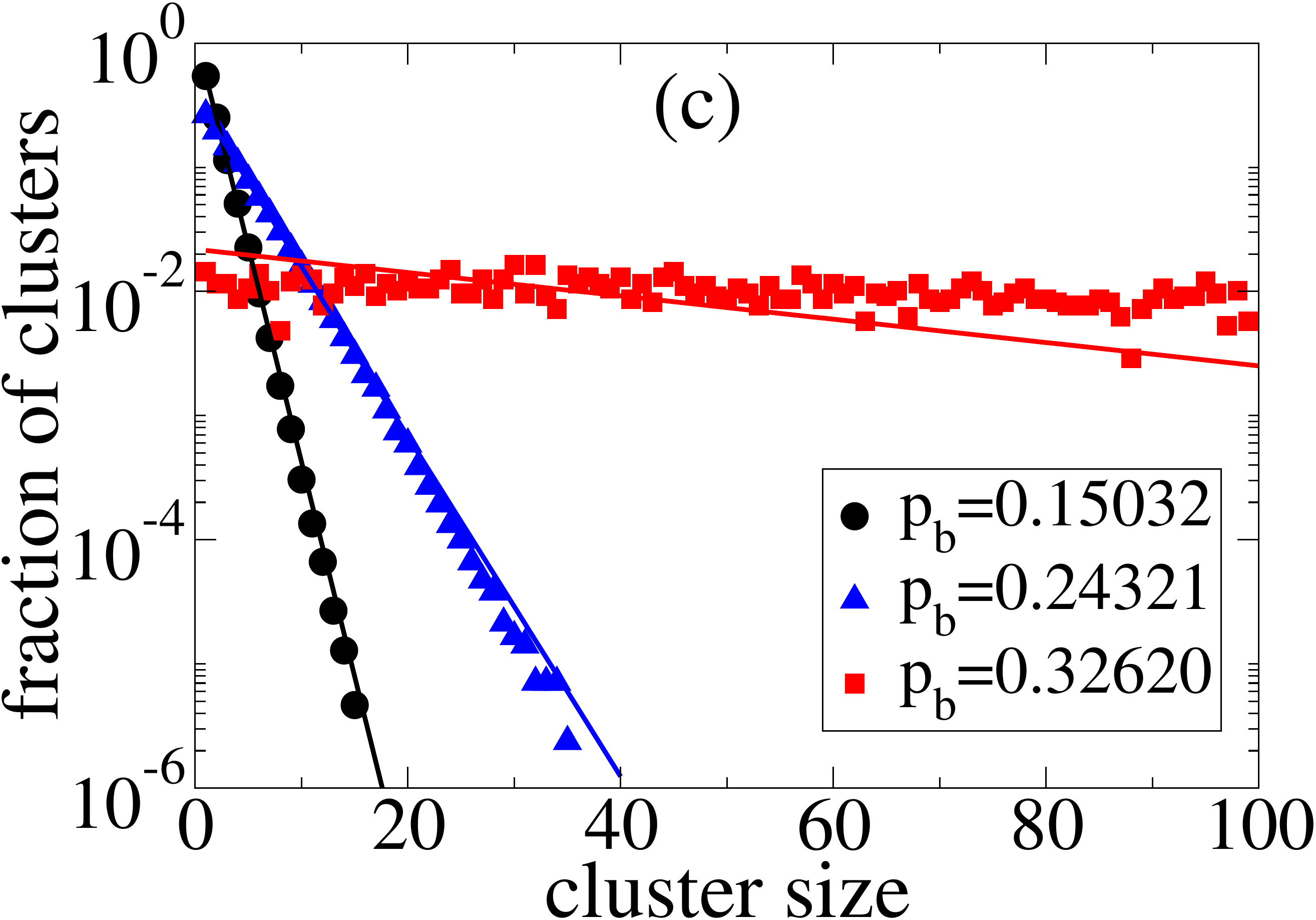}
\includegraphics[width=0.45\columnwidth]{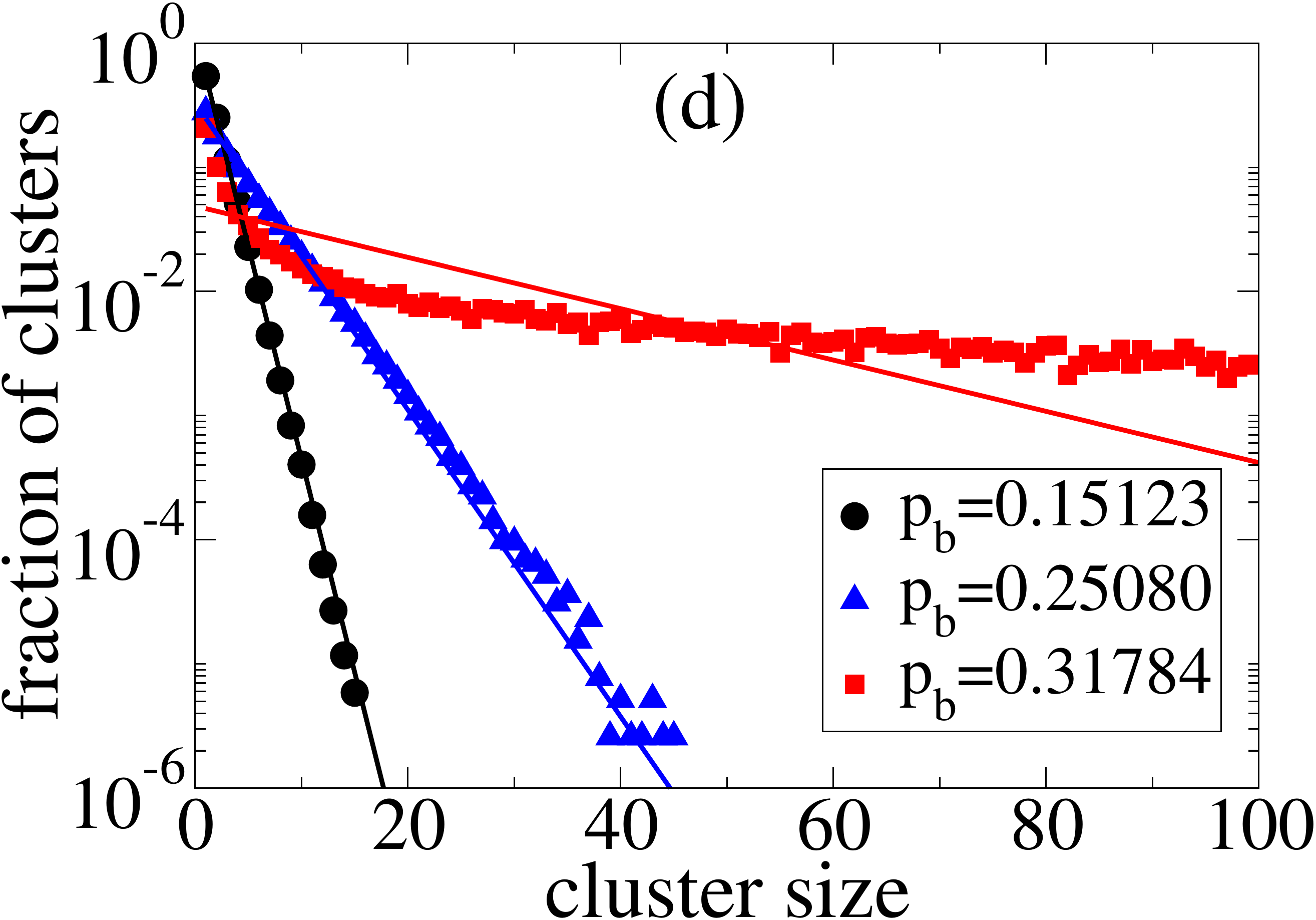}
%}
\caption{Fraction of the clusters of a given size for $\Delta=10, f=6$ at different times (corresponding to the values of $p_b$). (a) $\phi=0.05$; (b) $\phi=0.95$; (c) $\phi=0.5$, (d) $\phi=0.2$. Symbols are the simulation results and the lines are obtained from Eq. (\ref{fracclusti}) at the indicated values of $p_b$. In all cases, lines with larger (absolute) slopes correspond to larger values of $p_b$. ~\label{fig.9} }
\end{figure}

The fraction of clusters of a given size $i$, $m_i/M_0$, is represented in Fig. \ref{fig.9} at several values of $\phi$ and three different times, corresponding to $p_b\approx 0.5 p_b(\infty)$, $p_b\approx 0.75 p_b(\infty)$ and $p_b\approx p_b(\infty)$. The theoretical results are obtained using Eq. (\ref{finalmi}) for the fraction of clusters of size $i$,
\begin{equation}
\label{fracclusti}
\frac{m_i}{M_0}=\left(1-\frac{f p_b}{2}\right)\left(\frac{fp_b}{2}\right)^{i-1},
\end{equation}
where $p_b=2\phi p_2$.  
At the initial stage of growth,  the cluster size distributions are exponential for every $\phi$ and the agreement between Eq. (\ref{fracclusti}) and the results of simulation is excellent.  As time progresses, we find that this distribution is still exponential for large $i$, at $\phi=0.05$ and $\phi=0.95$ (see Figs. \ref{fig.9}(a) and (b)), but the agreement between theory and simulations is poor. The theory predicts an exponential distribution for all sizes and largely underestimates the frequency of large clusters. These differences are compatible with the results reported in Figs. \ref{fig.8}(a) and (b): theory and simulation agree for the number of small clusters and for $\langle S_1 \rangle$ (which is dominated by the contribution of the more abundant small clusters); on the other hand, as large clusters have a more significant contribution to $M_2$, the broader distributions observed in the simulations yield a larger $M_2$ than that predicted by the theory.

There is a remarkable agreement between the cluster size distributions obtained by simulations and theory at $\phi=0.5$ (see Fig. \ref{fig.9}(c)). At this value of $\phi$ the asymptotic value of $p_b$ is $2/f$, and thus the distribution Eq. (\ref{fracclusti}) becomes a very slow varying function of $i$ for values of $p_b$ close to $2/f$.  
%\cite{redner},
%\begin{equation}
%\label{fracclustass}
%\frac{m_i}{M_0}\approx A \exp\left(-iA\right),
%\end{equation}
%with $A=2(1-f p_b/2)/(fp_b)$.
Both theory and simulation exhibit a very broad cluster size distribution (almost uniform in the simulations) at the highest value of $p_b$ in Fig. \ref{fig.9}(c). To test if this agreement is obtained at other values of $\phi \in [1/f;1-1/f]$, the results at $\phi=0.2$ are shown in Fig. \ref{fig.9}(d). Exponential distributions are obtained at low and intermediate values of $p_b$ and for large cluster sizes only when $p_b$ is close to $2/f$. The theory understimates both the number of very small and very large clusters. 
Still, 
the broadness of $m_i/M_0$ is reached asymptotically in both
theory and simulations.

In summary, both the theory and the simulations predict two different behaviours for the cluster size distribution: (i) at high and low $\phi$, the distribution will be dominated by very small clusters and the growth of large clusters is rather limited; (ii) at intermediate values of $\phi$ the distribution will evolve to almost uniform, with a large number of small and large clusters alike.

\subsection{Comparison with experiments}

The theoretical framework developed here  can be tested by  comparison with  experiments reported in \cite{Ghofraniha2009}, where the assembly kinetics of  dilute binary mixtures  of streptavidin-coated large 
particles (P) and biotin-coated small ones (L - the linkers) \cite{Hiddessen2000} was studied using Dynamic Light Scattering\footnote{By an unfortunate coincidence the strepatividin-coated particles are labeled in 
\cite{Ghofraniha2009} 
%$\left[1\right]$ 
with L (for large) and the 
biotin-coated particles with S (for small). We made the option of keeping our notation for particles and linkers. Please be aware of this when contrasting our results to those of 
\cite{Ghofraniha2009}}.
%$\left[1\right]$}. 
From these measurements the time evolution of the number of clusters  and  the number of linkers (i.e. $m_i(t)$ for different values of $i$ and $N_0(t)$) was obtained. The number of particles, $N_P$, was fixed and  samples   with  different number of linkers, $N_L$, were investigated. A combination of geometrical parameters with the analysis of experimental results  led to the estimate $f=65$ as the maximum number of linkers that can bond to a particle. The diffusion coefficients of particles and linkers were measured in an independent experiment and $D_L/D_P\approx 5$ was found.

\begin{figure} [tb]
\vspace{0.5cm}
\includegraphics[width=\columnwidth]{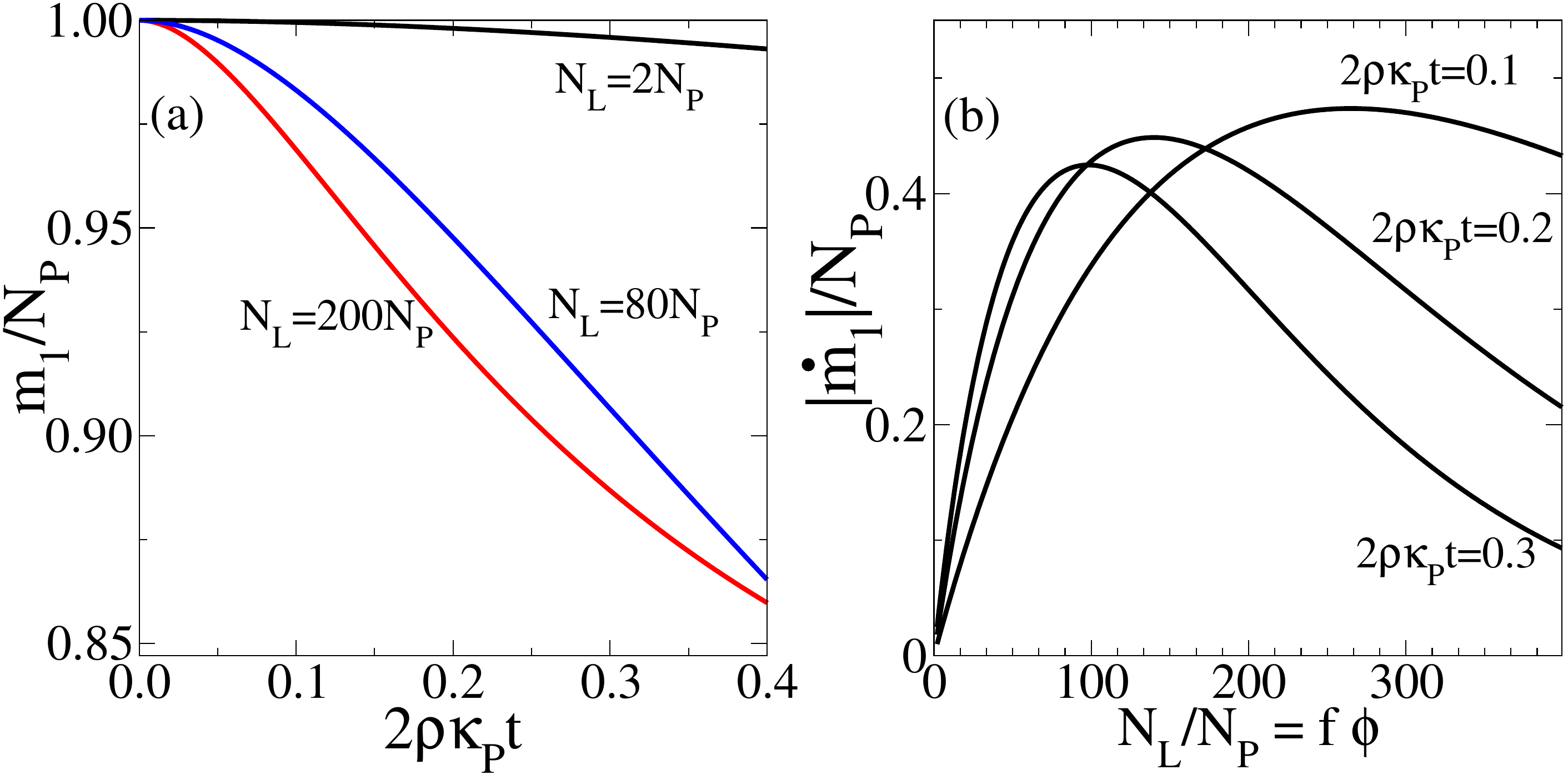}
\caption{ (a) Time evolution of the number of monomeric particles $m_1$ as predicted by  theory for three values of $N_L/N_P$ used in the experiments; (b) Theoretical prediction for the rate of change  of the number of monomeric particles $\dot m_1$ at three different times as a function of $N_L/N_P$. The model parameters used in the calculations are those reported in the experiments, $f=65$ and $\Delta=5$. This figure reproduces the experimental results reported in figure 5 of \cite{Ghofraniha2009}. ~\label{expfig.1}}
\end{figure}

%In the following, we show that the theory developed in the previous sections can predict and explain some results of \cite{sciortino2009} in the 
The analysis of the experimental results for $m_1(t)$ (figure 5 of \cite{Ghofraniha2009}) shows that, in the inital stages of growth, the rate at which monomeric particles disappear has a non-monotonic dependence on $N_L$: at a given instant, $|\dot m_1|$ was found to be maximal  for an intermediate value of $N_L$.
The prediction of our theory for $m_1(t)$ is obtained by solving Eq. (\ref{dotmi}) for $i=1$,
\begin{equation}
\label{dotm1}
\dot m_1=-2m_1q(1-q)(1-f\phi p_2),
\end{equation}
together with Eqs. (\ref{dotp0}) and (\ref{dotp2}), using the experimental parameters $\Delta=5$ (and $R_L/R_P=1/\Delta$ in (\ref{dotp0})), $f=65$, and $f\phi=N_L/N_P$. Figure \ref{expfig.1}(a) shows the results of this calculation  for the experimental values $N_L/N_P=2, 80$, and $200$: the faster variation is obtained first for $N_L=200$ and then for $N_L=80$. 
This non trivial behaviour is elucidated in Fig. \ref{expfig.1}(b):   $|\dot m_1|$ is represented as a function of $N_L/N_P$ at three different times in the initial stages of growth ($2\rho\kappa_P t=0.1, 0.2$ and $0.3$) and is in line with the non-monotonic experimental observation depicted in figure 5(b) of Ref. \cite{Ghofraniha2009}. 
%The presence of a large number $N_L>>fN_P$ of linkers at $t=0$ promotes the fast initial formation of linker particle bonds and, as a contribution to %$\dot m_1(t)$, of some dimers. However, in a time that is smaller the larger the value of $N_L$ is,  dimer formation is hampered due to the %presence of more and more particles that are fully covered with linkers (since $N_L\gg f$) and  $\dot m_1$ decreases. This can be clearly seen in %figure 
%\ref{expfig.1}a for the case $N_L=200$. If the number of 

Notice that the numerical simulations reported in \cite{Ghofraniha2009} were not able to reproduce the behaviour depicted in Fig. \ref{expfig.1}, since they were limited by technical reasons to extremely short times.  By contrast, in the  theory described here, all the parameters (time included) can be varied freely,  providing a simple tool for interpreting and predicting  experimental results.

\begin{figure} [tbh]
\includegraphics[width=\columnwidth]{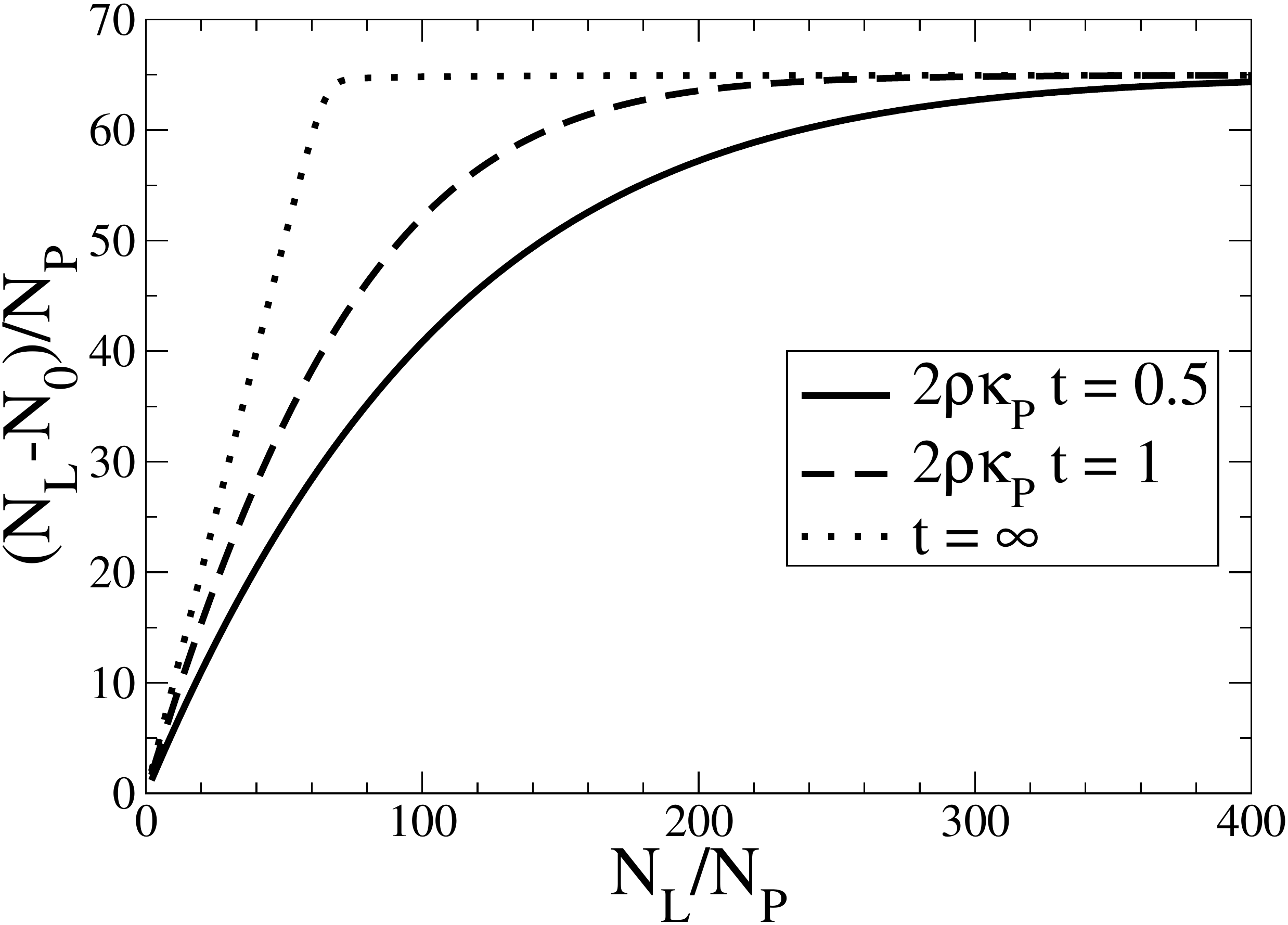}
\caption{Theoretical prediction for the number of bonded linkers per particle  at the two indicated (long) times as a function of $N_L/N_P$.   The dotted line represents the asymptotic limit and corresponds, for each $N_L/N_P$, to full adsorption of linkers on particles. $\Delta=5$ and $f=65$ were used in the calculations. This figure reproduces the experimental results reported in figure 3 of \cite{Ghofraniha2009}.~\label{expfig.2}}
\end{figure}

In particular the theory describes the long time behaviour reported in \cite{Ghofraniha2009}.
The asymptotic regime (limit $t\to \infty$) of the kinetics was analysed  by  calculating the fraction of linkers bonded to particles (using the measurement of the number of free linkers $N_0(t)$ at long  observation time) - see figure 3 of \cite{Ghofraniha2009}. 
 The observation that  the  full adsorption of linkers by particles only occurs at high values of 
$N_L/N_P$ is reported as unexpected.
  In Fig. \ref{expfig.2} we 
plot the number of linkers bonded per particle, i.e. $(N_L-N_0(t))/N_P=f\phi(1-p_0(t))$ , as a function of $N_L/N_P$ at two instants, calculated using Eqs. (\ref{dotp0}) and (\ref{dotp2}). We also plot the asymptotic limit of the theory that indeed corresponds to the full adsorption case. However, it is clear that the time to reach this limit is strongly dependent on $N_L/N_P$. It is possible to observe, at the same time, samples at  full adsorption and samples that are only halfway   to that limit  (e.g. $N_L/N_P\approx 400$ and $N_L/N_P\approx 70$, respectively, for $2\rho\kappa_P t=0.5$). 

% The purpose of these comparisons is to illustrate the applicability to real systems of the relatively simple theory that we have proposed in this paper. %A  full analysis of the experimental results would most probably lead to a better

\section{Conclusions and discussion} \label{sec.conclusion}
We have studied 
%{\xout{a simple model of}} 
 the dynamics of linker-mediated irreversible aggregation of particles with a maximum valence  $f$, by  developing a generalized Smoluchowski theoretical framework 
in which linkers and particles are explicitly taken into account. 
%{\xout{This process was analysed using simulations and a simplified version of the  Smoluchowski equations.}}.  
Smoluchowski equations were solved using simulations and an approximated theory obtained via a scaling argument.  
Simulations have shown that the dynamics and the cluster size distributions are determined by $\phi$ (the ratio between the actual number of linkers and the maximum number of linkers to which particles can bond or total number of bonding sites) and by $\Delta$ (the ratio between the diffusion coefficients of linkers and clusters)  in a non-trivial way. In particular, $\phi$ alone determines the asymptotic state of the system when $\phi<1-1/f$, while for
% {\xout{$\phi>1-1/f$}} 
 $1>\phi>1-1/f$  (i.e. for a large number of linkers) the asymptotic state will depend also on $\Delta$. The cluster size distributions obtained at both low ($<1/f$) and high ($>1-1/f$) $\phi$ are exponential and exhibit finite and small moments at all times. On the other hand, for $\phi \in [1/f;1-1/f]$ the distribution becomes very broad asymptotically,  with a diverging second moment.  In this interval, the initial growth of the clusters is faster at larger values of $\phi$ and $\Delta$; however, after a crossover time that depends on $\Delta$, the system with $\phi=0.5$ exhibits always the fastest growth and the largest clusters (at a given time). 

The results of the  approximate analytical 
%{\xout{Smoluchowski equations (\ref{dotmi},\ref{dotN01})} 
theory -Eqs. (\ref{dotp0},\ref{dotp2})- are in line qualitatively (and in some cases quantitavely)  with the simulation results. This comparison tests the validity of the assumption that the fraction of unnocupied sites of a cluster is independent of its size. This hypothesis works quantitatively at the initial stages of growth but fails drastically at low ($<1/f$) and high ($>1-1/f$) $\phi$ by failing to describe correctly the cluster size distributions at long times. This means that, at least at low and high $\phi$, there is a relevant non trivial time evolution of the dependence on the cluster size of the fraction of unoccupied sites that is not captured by  the analytical theory.
The theory makes the strong prediction that the cluster size distribution (\ref {fracclusti}) 
depends on time only through the bonding probability $p_b$, and the comparison with the simulations shows that this is valid for values of $\phi \in 
[1/f;1-1/f]$.   Any extension of this theory would  have to include a explicit $\phi$ and cluster size dependence on the fraction of occupied sites. 
%of a cluster.

The cluster size distribution (\ref{fracclusti}) of the linker-mediated aggregation is equal to that obtained for a constant kernel \cite{Krapivsky2011},  and therefore the model proposed in this work, with dynamics determined by 
%{\xout{a ``normalized''  polymerization kernel (\ref{pi1j1i2j2})}
the kernel defined in Eqs. (\ref{kernel1ap},\ref{kernel0ap}), turns out to be similar to the model of irreversible aggregation with a constant kernel. In particular, the absence of percolation (i.e. the formation of a giant cluster in a finite time) and the asymptotic broad distribution at large times reinforce the similarity with the linker-mediated aggregation model in the range $\phi \in [1/f;1-1/f]$.  

 We  found that the analytical theory explains some  experimental results of \cite{Ghofraniha2009}. In particular, the simplicity of the theory allowed to vary both $f\phi=N_L/N_P$ and time, enabling the  exploration of  length and time scales that are not accessible through detailed numerical simulations.

We focused on  the study of limited linker-particle aggregation with two diffusion time scales. 
  One could extend this study  by adding more realistic ingredients  to the theory
   such as:
  (i) clusters with more general structures (not just tree like), and even loop formation, since it seems to be particularly important, at least in equilibrium conditions, when the linkers are polymer like \cite{Howard2019}; (ii) other kernels, namely a generalized polymerization kernel, which may be  amenable to  analytical treatment and compared with the approximation developed in \cite{Sciortino2009}; (iii) correlations in the distribution of  unoccupied  sites, etc. However, our study shows that this type of theories, regardless of the strong simplifying assumptions, do contribute to a deeper  understanding of the dynamics of linker-mediated aggregation.

\section{acknowledgments} We acknowledge financial support from the Portuguese
Foundation for Science and Technology (FCT) under Contracts no.
PTDC/FIS-MAC/28146/2017 (LISBOA-01-0145-FEDER-028146), UIDB/00618/2020,
UIDP/00618/2020, and CEECIND/00586/2017.

%\begin{thebibliography}*
%\bibliography{library}
\bibliography{linkers}

\begin{thebibliography}{27}%
\makeatletter
\providecommand \@ifxundefined [1]{%
 \@ifx{#1\undefined}
}%
\providecommand \@ifnum [1]{%
 \ifnum #1\expandafter \@firstoftwo
 \else \expandafter \@secondoftwo
 \fi
}%
\providecommand \@ifx [1]{%
 \ifx #1\expandafter \@firstoftwo
 \else \expandafter \@secondoftwo
 \fi
}%
\providecommand \natexlab [1]{#1}%
\providecommand \enquote  [1]{``#1''}%
\providecommand \bibnamefont  [1]{#1}%
\providecommand \bibfnamefont [1]{#1}%
\providecommand \citenamefont [1]{#1}%
\providecommand \href@noop [0]{\@secondoftwo}%
\providecommand \href [0]{\begingroup \@sanitize@url \@href}%
\providecommand \@href[1]{\@@startlink{#1}\@@href}%
\providecommand \@@href[1]{\endgroup#1\@@endlink}%
\providecommand \@sanitize@url [0]{\catcode `\\12\catcode `\$12\catcode
  `\&12\catcode `\#12\catcode `\^12\catcode `\_12\catcode `\%12\relax}%
\providecommand \@@startlink[1]{}%
\providecommand \@@endlink[0]{}%
\providecommand \url  [0]{\begingroup\@sanitize@url \@url }%
\providecommand \@url [1]{\endgroup\@href {#1}{\urlprefix }}%
\providecommand \urlprefix  [0]{URL }%
\providecommand \Eprint [0]{\href }%
\providecommand \doibase [0]{http://dx.doi.org/}%
\providecommand \selectlanguage [0]{\@gobble}%
\providecommand \bibinfo  [0]{\@secondoftwo}%
\providecommand \bibfield  [0]{\@secondoftwo}%
\providecommand \translation [1]{[#1]}%
\providecommand \BibitemOpen [0]{}%
\providecommand \bibitemStop [0]{}%
\providecommand \bibitemNoStop [0]{.\EOS\space}%
\providecommand \EOS [0]{\spacefactor3000\relax}%
\providecommand \BibitemShut  [1]{\csname bibitem#1\endcsname}%
\let\auto@bib@innerbib\@empty
%</preamble>
\bibitem [{\citenamefont {Hiddessen}\ \emph {et~al.}(2000)\citenamefont
  {Hiddessen}, \citenamefont {Rodgers}, \citenamefont {Weitz},\ and\
  \citenamefont {Hammer}}]{Hiddessen2000}%
  \BibitemOpen
  \bibfield  {author} {\bibinfo {author} {\bibfnamefont {A.~L.}\ \bibnamefont
  {Hiddessen}}, \bibinfo {author} {\bibfnamefont {S.~D.}\ \bibnamefont
  {Rodgers}}, \bibinfo {author} {\bibfnamefont {D.~A.}\ \bibnamefont {Weitz}},
  \ and\ \bibinfo {author} {\bibfnamefont {D.~A.}\ \bibnamefont {Hammer}},\
  }\href@noop {} {\bibfield  {journal} {\bibinfo  {journal} {Langmuir}\
  }\textbf {\bibinfo {volume} {16}},\ \bibinfo {pages} {9744} (\bibinfo {year}
  {2000})}\BibitemShut {NoStop}%
\bibitem [{\citenamefont {Milam}\ \emph {et~al.}(2003)\citenamefont {Milam},
  \citenamefont {Hiddessen}, \citenamefont {Crocker}, \citenamefont {Graves},\
  and\ \citenamefont {Hammer}}]{Milam2003}%
  \BibitemOpen
  \bibfield  {author} {\bibinfo {author} {\bibfnamefont {V.~T.}\ \bibnamefont
  {Milam}}, \bibinfo {author} {\bibfnamefont {A.~L.}\ \bibnamefont
  {Hiddessen}}, \bibinfo {author} {\bibfnamefont {J.~C.}\ \bibnamefont
  {Crocker}}, \bibinfo {author} {\bibfnamefont {D.~J.}\ \bibnamefont {Graves}},
  \ and\ \bibinfo {author} {\bibfnamefont {D.~A.}\ \bibnamefont {Hammer}},\
  }\href@noop {} {\bibfield  {journal} {\bibinfo  {journal} {Langmuir}\
  }\textbf {\bibinfo {volume} {19}},\ \bibinfo {pages} {10317} (\bibinfo {year}
  {2003})}\BibitemShut {NoStop}%
\bibitem [{\citenamefont {Hiddessen}\ \emph {et~al.}(2004)\citenamefont
  {Hiddessen}, \citenamefont {Weitz},\ and\ \citenamefont
  {Hammer}}]{Hiddessen2004}%
  \BibitemOpen
  \bibfield  {author} {\bibinfo {author} {\bibfnamefont {A.~L.}\ \bibnamefont
  {Hiddessen}}, \bibinfo {author} {\bibfnamefont {D.~A.}\ \bibnamefont
  {Weitz}}, \ and\ \bibinfo {author} {\bibfnamefont {D.~A.}\ \bibnamefont
  {Hammer}},\ }\href@noop {} {\bibfield  {journal} {\bibinfo  {journal}
  {Langmuir}\ }\textbf {\bibinfo {volume} {20}},\ \bibinfo {pages} {6788}
  (\bibinfo {year} {2004})}\BibitemShut {NoStop}%
\bibitem [{\citenamefont {Ghofraniha}\ \emph {et~al.}(2009)\citenamefont
  {Ghofraniha}, \citenamefont {Andreozzi}, \citenamefont {Russo}, \citenamefont
  {{La Mesa}},\ and\ \citenamefont {Sciortino}}]{Ghofraniha2009}%
  \BibitemOpen
  \bibfield  {author} {\bibinfo {author} {\bibfnamefont {N.}~\bibnamefont
  {Ghofraniha}}, \bibinfo {author} {\bibfnamefont {P.}~\bibnamefont
  {Andreozzi}}, \bibinfo {author} {\bibfnamefont {J.}~\bibnamefont {Russo}},
  \bibinfo {author} {\bibfnamefont {C.}~\bibnamefont {{La Mesa}}}, \ and\
  \bibinfo {author} {\bibfnamefont {F.}~\bibnamefont {Sciortino}},\ }\href@noop
  {} {\bibfield  {journal} {\bibinfo  {journal} {J Phys. Chem. B}\ }\textbf
  {\bibinfo {volume} {113}},\ \bibinfo {pages} {6775} (\bibinfo {year}
  {2009})}\BibitemShut {NoStop}%
\bibitem [{\citenamefont {Bharti}\ \emph {et~al.}(2014)\citenamefont {Bharti},
  \citenamefont {Meissner}, \citenamefont {Klapp},\ and\ \citenamefont
  {Findenegg}}]{Bharti2014}%
  \BibitemOpen
  \bibfield  {author} {\bibinfo {author} {\bibfnamefont {B.}~\bibnamefont
  {Bharti}}, \bibinfo {author} {\bibfnamefont {J.}~\bibnamefont {Meissner}},
  \bibinfo {author} {\bibfnamefont {S.~H.~L.}\ \bibnamefont {Klapp}}, \ and\
  \bibinfo {author} {\bibfnamefont {G.~H.}\ \bibnamefont {Findenegg}},\
  }\href@noop {} {\bibfield  {journal} {\bibinfo  {journal} {Soft Matt.}\
  }\textbf {\bibinfo {volume} {10}},\ \bibinfo {pages} {718} (\bibinfo {year}
  {2014})}\BibitemShut {NoStop}%
\bibitem [{\citenamefont {Peng}\ \emph {et~al.}(2016)\citenamefont {Peng},
  \citenamefont {Kroes-Nijboer}, \citenamefont {Venema},\ and\ \citenamefont
  {van~der Linden}}]{Peng2016}%
  \BibitemOpen
  \bibfield  {author} {\bibinfo {author} {\bibfnamefont {J.}~\bibnamefont
  {Peng}}, \bibinfo {author} {\bibfnamefont {A.}~\bibnamefont {Kroes-Nijboer}},
  \bibinfo {author} {\bibfnamefont {P.}~\bibnamefont {Venema}}, \ and\ \bibinfo
  {author} {\bibfnamefont {E.}~\bibnamefont {van~der Linden}},\ }\href@noop {}
  {\bibfield  {journal} {\bibinfo  {journal} {Soft Matt.}\ }\textbf {\bibinfo
  {volume} {12}},\ \bibinfo {pages} {3514} (\bibinfo {year}
  {2016})}\BibitemShut {NoStop}%
\bibitem [{\citenamefont {Singh}\ \emph {et~al.}(2015)\citenamefont {Singh},
  \citenamefont {Lindquist}, \citenamefont {Ong}, \citenamefont {Jadrich},
  \citenamefont {Singh}, \citenamefont {Ha}, \citenamefont {Ellison},
  \citenamefont {Truskett},\ and\ \citenamefont {Milliron}}]{Singh2015}%
  \BibitemOpen
  \bibfield  {author} {\bibinfo {author} {\bibfnamefont {A.}~\bibnamefont
  {Singh}}, \bibinfo {author} {\bibfnamefont {B.~A.}\ \bibnamefont
  {Lindquist}}, \bibinfo {author} {\bibfnamefont {G.~K.}\ \bibnamefont {Ong}},
  \bibinfo {author} {\bibfnamefont {R.~B.}\ \bibnamefont {Jadrich}}, \bibinfo
  {author} {\bibfnamefont {A.}~\bibnamefont {Singh}}, \bibinfo {author}
  {\bibfnamefont {H.}~\bibnamefont {Ha}}, \bibinfo {author} {\bibfnamefont
  {C.~J.}\ \bibnamefont {Ellison}}, \bibinfo {author} {\bibfnamefont {T.~M.}\
  \bibnamefont {Truskett}}, \ and\ \bibinfo {author} {\bibfnamefont {D.~J.}\
  \bibnamefont {Milliron}},\ }\href@noop {} {\bibfield  {journal} {\bibinfo
  {journal} {Angew. Chem. Int. Ed.}\ }\textbf {\bibinfo {volume} {54}},\
  \bibinfo {pages} {14840} (\bibinfo {year} {2015})}\BibitemShut {NoStop}%
\bibitem [{\citenamefont {Fernandez-Castanon}\ \emph
  {et~al.}(2018)\citenamefont {Fernandez-Castanon}, \citenamefont {Bomboi},\
  and\ \citenamefont {Sciortino}}]{Castanon2018}%
  \BibitemOpen
  \bibfield  {author} {\bibinfo {author} {\bibfnamefont {J.}~\bibnamefont
  {Fernandez-Castanon}}, \bibinfo {author} {\bibfnamefont {F.}~\bibnamefont
  {Bomboi}}, \ and\ \bibinfo {author} {\bibfnamefont {F.}~\bibnamefont
  {Sciortino}},\ }\href@noop {} {\bibfield  {journal} {\bibinfo  {journal} {J.
  Chem. Phys.}\ }\textbf {\bibinfo {volume} {148}},\ \bibinfo {pages} {025103}
  (\bibinfo {year} {2018})}\BibitemShut {NoStop}%
\bibitem [{\citenamefont {Lowensohn}\ \emph {et~al.}(2019)\citenamefont
  {Lowensohn}, \citenamefont {Oyarz{\'{u}}n}, \citenamefont {{Narv{\'{a}}ez
  Paliza}}, \citenamefont {Mognetti},\ and\ \citenamefont
  {Rogers}}]{Lowensohn2019}%
  \BibitemOpen
  \bibfield  {author} {\bibinfo {author} {\bibfnamefont {J.}~\bibnamefont
  {Lowensohn}}, \bibinfo {author} {\bibfnamefont {B.}~\bibnamefont
  {Oyarz{\'{u}}n}}, \bibinfo {author} {\bibfnamefont {G.}~\bibnamefont
  {{Narv{\'{a}}ez Paliza}}}, \bibinfo {author} {\bibfnamefont {B.~M.}\
  \bibnamefont {Mognetti}}, \ and\ \bibinfo {author} {\bibfnamefont {W.~B.}\
  \bibnamefont {Rogers}},\ }\href@noop {} {\bibfield  {journal} {\bibinfo
  {journal} {Phys. Rev. X}\ }\textbf {\bibinfo {volume} {9}},\ \bibinfo {pages}
  {41054} (\bibinfo {year} {2019})}\BibitemShut {NoStop}%
\bibitem [{\citenamefont {Antunes}\ \emph {et~al.}(2019)\citenamefont
  {Antunes}, \citenamefont {Dias}, \citenamefont {{Telo Da Gama}},\ and\
  \citenamefont {Ara{\'{u}}jo}}]{Antunes2019}%
  \BibitemOpen
  \bibfield  {author} {\bibinfo {author} {\bibfnamefont {G.~C.}\ \bibnamefont
  {Antunes}}, \bibinfo {author} {\bibfnamefont {C.~S.}\ \bibnamefont {Dias}},
  \bibinfo {author} {\bibfnamefont {M.~M.}\ \bibnamefont {{Telo Da Gama}}}, \
  and\ \bibinfo {author} {\bibfnamefont {N.~A.~M.}\ \bibnamefont
  {Ara{\'{u}}jo}},\ }\href@noop {} {\bibfield  {journal} {\bibinfo  {journal}
  {Soft Matt.}\ }\textbf {\bibinfo {volume} {15}},\ \bibinfo {pages} {3712}
  (\bibinfo {year} {2019})}\BibitemShut {NoStop}%
\bibitem [{\citenamefont {Sciortino}\ \emph {et~al.}(2009)\citenamefont
  {Sciortino}, \citenamefont {{De Michele}}, \citenamefont {Corezzi},
  \citenamefont {Russo}, \citenamefont {Zaccarelli},\ and\ \citenamefont
  {Tartaglia}}]{Sciortino2009}%
  \BibitemOpen
  \bibfield  {author} {\bibinfo {author} {\bibfnamefont {F.}~\bibnamefont
  {Sciortino}}, \bibinfo {author} {\bibfnamefont {C.}~\bibnamefont {{De
  Michele}}}, \bibinfo {author} {\bibfnamefont {S.}~\bibnamefont {Corezzi}},
  \bibinfo {author} {\bibfnamefont {J.}~\bibnamefont {Russo}}, \bibinfo
  {author} {\bibfnamefont {E.}~\bibnamefont {Zaccarelli}}, \ and\ \bibinfo
  {author} {\bibfnamefont {P.}~\bibnamefont {Tartaglia}},\ }\href@noop {}
  {\bibfield  {journal} {\bibinfo  {journal} {Soft Matt.}\ }\textbf {\bibinfo
  {volume} {5}},\ \bibinfo {pages} {2571} (\bibinfo {year} {2009})}\BibitemShut
  {NoStop}%
\bibitem [{\citenamefont {Lindquist}\ \emph {et~al.}(2016)\citenamefont
  {Lindquist}, \citenamefont {Jadrich}, \citenamefont {Milliron},\ and\
  \citenamefont {Truskett}}]{Lindquist2016}%
  \BibitemOpen
  \bibfield  {author} {\bibinfo {author} {\bibfnamefont {B.~A.}\ \bibnamefont
  {Lindquist}}, \bibinfo {author} {\bibfnamefont {R.~B.}\ \bibnamefont
  {Jadrich}}, \bibinfo {author} {\bibfnamefont {D.~J.}\ \bibnamefont
  {Milliron}}, \ and\ \bibinfo {author} {\bibfnamefont {T.~M.}\ \bibnamefont
  {Truskett}},\ }\href@noop {} {\bibfield  {journal} {\bibinfo  {journal} {J.
  Chem. Phys.}\ }\textbf {\bibinfo {volume} {145}},\ \bibinfo {pages} {074906}
  (\bibinfo {year} {2016})}\BibitemShut {NoStop}%
\bibitem [{\citenamefont {Corezzi}\ \emph {et~al.}(2010)\citenamefont
  {Corezzi}, \citenamefont {Fioretto}, \citenamefont {{De Michele}},
  \citenamefont {Zaccarelli},\ and\ \citenamefont {Sciortino}}]{Corezzi2010}%
  \BibitemOpen
  \bibfield  {author} {\bibinfo {author} {\bibfnamefont {S.}~\bibnamefont
  {Corezzi}}, \bibinfo {author} {\bibfnamefont {D.}~\bibnamefont {Fioretto}},
  \bibinfo {author} {\bibfnamefont {C.}~\bibnamefont {{De Michele}}}, \bibinfo
  {author} {\bibfnamefont {E.}~\bibnamefont {Zaccarelli}}, \ and\ \bibinfo
  {author} {\bibfnamefont {F.}~\bibnamefont {Sciortino}},\ }\href@noop {}
  {\bibfield  {journal} {\bibinfo  {journal} {J. Phys. Chem. B}\ }\textbf
  {\bibinfo {volume} {114}},\ \bibinfo {pages} {3769} (\bibinfo {year}
  {2010})}\BibitemShut {NoStop}%
\bibitem [{\citenamefont {Cyron}\ \emph {et~al.}(2013)\citenamefont {Cyron},
  \citenamefont {M{\"{u}}ller}, \citenamefont {Schmoller}, \citenamefont
  {Bausch}, \citenamefont {Wall},\ and\ \citenamefont {Bruinsma}}]{Cyron2013}%
  \BibitemOpen
  \bibfield  {author} {\bibinfo {author} {\bibfnamefont {C.~J.}\ \bibnamefont
  {Cyron}}, \bibinfo {author} {\bibfnamefont {K.~W.}\ \bibnamefont
  {M{\"{u}}ller}}, \bibinfo {author} {\bibfnamefont {K.~M.}\ \bibnamefont
  {Schmoller}}, \bibinfo {author} {\bibfnamefont {A.~R.}\ \bibnamefont
  {Bausch}}, \bibinfo {author} {\bibfnamefont {W.~A.}\ \bibnamefont {Wall}}, \
  and\ \bibinfo {author} {\bibfnamefont {R.~F.}\ \bibnamefont {Bruinsma}},\
  }\href@noop {} {\bibfield  {journal} {\bibinfo  {journal} {EPL}\ }\textbf
  {\bibinfo {volume} {102}},\ \bibinfo {pages} {38003} (\bibinfo {year}
  {2013})}\BibitemShut {NoStop}%
\bibitem [{\citenamefont {Pierce}\ \emph {et~al.}(2004)\citenamefont {Pierce},
  \citenamefont {Chakrabarti}, \citenamefont {Fry},\ and\ \citenamefont
  {Sorensen}}]{Pierce2004}%
  \BibitemOpen
  \bibfield  {author} {\bibinfo {author} {\bibfnamefont {F.}~\bibnamefont
  {Pierce}}, \bibinfo {author} {\bibfnamefont {A.}~\bibnamefont {Chakrabarti}},
  \bibinfo {author} {\bibfnamefont {D.}~\bibnamefont {Fry}}, \ and\ \bibinfo
  {author} {\bibfnamefont {C.~M.}\ \bibnamefont {Sorensen}},\ }\href@noop {}
  {\bibfield  {journal} {\bibinfo  {journal} {Langmuir}\ }\textbf {\bibinfo
  {volume} {20}},\ \bibinfo {pages} {2498} (\bibinfo {year}
  {2004})}\BibitemShut {NoStop}%
\bibitem [{\citenamefont {Wang}\ \emph {et~al.}(2020)\citenamefont {Wang},
  \citenamefont {Lee},\ and\ \citenamefont {Arya}}]{Wang2020a}%
  \BibitemOpen
  \bibfield  {author} {\bibinfo {author} {\bibfnamefont {J.}~\bibnamefont
  {Wang}}, \bibinfo {author} {\bibfnamefont {B.~H.~J.}\ \bibnamefont {Lee}}, \
  and\ \bibinfo {author} {\bibfnamefont {G.}~\bibnamefont {Arya}},\ }\href@noop
  {} {\bibfield  {journal} {\bibinfo  {journal} {Nanoscale}\ }\textbf {\bibinfo
  {volume} {12}},\ \bibinfo {pages} {5091} (\bibinfo {year}
  {2020})}\BibitemShut {NoStop}%
\bibitem [{\citenamefont {Xia}\ \emph {et~al.}(2020)\citenamefont {Xia},
  \citenamefont {Hu}, \citenamefont {Ciamarra},\ and\ \citenamefont
  {Ni}}]{Xia2020}%
  \BibitemOpen
  \bibfield  {author} {\bibinfo {author} {\bibfnamefont {X.}~\bibnamefont
  {Xia}}, \bibinfo {author} {\bibfnamefont {H.}~\bibnamefont {Hu}}, \bibinfo
  {author} {\bibfnamefont {M.~P.}\ \bibnamefont {Ciamarra}}, \ and\ \bibinfo
  {author} {\bibfnamefont {R.}~\bibnamefont {Ni}},\ }\href@noop {} {\bibfield
  {journal} {\bibinfo  {journal} {Sci. Adv.}\ }\textbf {\bibinfo {volume}
  {6}},\ \bibinfo {pages} {eaaz6921} (\bibinfo {year} {2020})}\BibitemShut
  {NoStop}%
\bibitem [{\citenamefont {M{\"{u}}ller}\ \emph {et~al.}(2014)\citenamefont
  {M{\"{u}}ller}, \citenamefont {Bruinsma}, \citenamefont {Lieleg},
  \citenamefont {Bausch}, \citenamefont {Wall},\ and\ \citenamefont
  {Levine}}]{Muller2014}%
  \BibitemOpen
  \bibfield  {author} {\bibinfo {author} {\bibfnamefont {K.~W.}\ \bibnamefont
  {M{\"{u}}ller}}, \bibinfo {author} {\bibfnamefont {R.~F.}\ \bibnamefont
  {Bruinsma}}, \bibinfo {author} {\bibfnamefont {O.}~\bibnamefont {Lieleg}},
  \bibinfo {author} {\bibfnamefont {A.~R.}\ \bibnamefont {Bausch}}, \bibinfo
  {author} {\bibfnamefont {W.~A.}\ \bibnamefont {Wall}}, \ and\ \bibinfo
  {author} {\bibfnamefont {A.~J.}\ \bibnamefont {Levine}},\ }\href@noop {}
  {\bibfield  {journal} {\bibinfo  {journal} {Phys. Rev. Lett.}\ }\textbf
  {\bibinfo {volume} {112}},\ \bibinfo {pages} {238102} (\bibinfo {year}
  {2014})}\BibitemShut {NoStop}%
\bibitem [{\citenamefont {Chandrasekhar}(1943)}]{Chandrasekhar43}%
  \BibitemOpen
  \bibfield  {author} {\bibinfo {author} {\bibfnamefont {S.}~\bibnamefont
  {Chandrasekhar}},\ }\href@noop {} {\bibfield  {journal} {\bibinfo  {journal}
  {Rev. Mod. Phys.}\ }\textbf {\bibinfo {volume} {15}},\ \bibinfo {pages} {1}
  (\bibinfo {year} {1943})}\BibitemShut {NoStop}%
\bibitem [{\citenamefont {Krapivsky}\ \emph {et~al.}(2011)\citenamefont
  {Krapivsky}, \citenamefont {Redner},\ and\ \citenamefont
  {Ben-Naim}}]{Krapivsky2011}%
  \BibitemOpen
  \bibfield  {author} {\bibinfo {author} {\bibfnamefont {P.~L.}\ \bibnamefont
  {Krapivsky}}, \bibinfo {author} {\bibfnamefont {S.}~\bibnamefont {Redner}}, \
  and\ \bibinfo {author} {\bibfnamefont {E.}~\bibnamefont {Ben-Naim}},\ }\href
  {\doibase 10.1017/CBO9780511780516} {\emph {\bibinfo {title} {A Kinetic View
  of Statistical Physics}}}\ (\bibinfo  {publisher} {Cambridge University
  Press, New York},\ \bibinfo {year} {2011})\ pp.\ \bibinfo {pages}
  {1--488}\BibitemShut {NoStop}%
\bibitem [{\citenamefont {van Dongen}\ and\ \citenamefont
  {Ernst}(1984)}]{VanDongen1984}%
  \BibitemOpen
  \bibfield  {author} {\bibinfo {author} {\bibfnamefont {P.~G.~J.}\
  \bibnamefont {van Dongen}}\ and\ \bibinfo {author} {\bibfnamefont {M.~H.}\
  \bibnamefont {Ernst}},\ }\href@noop {} {\bibfield  {journal} {\bibinfo
  {journal} {J. Stat. Phys.}\ }\textbf {\bibinfo {volume} {37}},\ \bibinfo
  {pages} {301} (\bibinfo {year} {1984})}\BibitemShut {NoStop}%
\bibitem [{\citenamefont {Aldous}(1999)}]{Aldous1999}%
  \BibitemOpen
  \bibfield  {author} {\bibinfo {author} {\bibfnamefont {D.~J.}\ \bibnamefont
  {Aldous}},\ }\href@noop {} {\bibfield  {journal} {\bibinfo  {journal}
  {Bernoulli}\ }\textbf {\bibinfo {volume} {5}},\ \bibinfo {pages} {3}
  (\bibinfo {year} {1999})}\BibitemShut {NoStop}%
\bibitem [{\citenamefont {Leyvraz}(1983)}]{Leyvraz1983}%
  \BibitemOpen
  \bibfield  {author} {\bibinfo {author} {\bibfnamefont {F.}~\bibnamefont
  {Leyvraz}},\ }\href@noop {} {\bibfield  {journal} {\bibinfo  {journal} {J.
  Phys. A: Math. Gen.}\ }\textbf {\bibinfo {volume} {16}},\ \bibinfo {pages}
  {2861} (\bibinfo {year} {1983})}\BibitemShut {NoStop}%
\bibitem [{\citenamefont {Smit}\ \emph {et~al.}(1994)\citenamefont {Smit},
  \citenamefont {Hounslow},\ and\ \citenamefont {Paterson}}]{Smit1994}%
  \BibitemOpen
  \bibfield  {author} {\bibinfo {author} {\bibfnamefont {D.~J.}\ \bibnamefont
  {Smit}}, \bibinfo {author} {\bibfnamefont {M.~J.}\ \bibnamefont {Hounslow}},
  \ and\ \bibinfo {author} {\bibfnamefont {W.~R.}\ \bibnamefont {Paterson}},\
  }\href@noop {} {\bibfield  {journal} {\bibinfo  {journal} {Chem. Eng. Sci.}\
  }\textbf {\bibinfo {volume} {49}},\ \bibinfo {pages} {1025} (\bibinfo {year}
  {1994})}\BibitemShut {NoStop}%
\bibitem [{SM()}]{SM}%
  \BibitemOpen
  \href {...} {}\bibinfo {note} {See Supplemental Material of ''Smoluchowski
  equations for linker-mediated irreversible aggregation''}\BibitemShut
  {NoStop}%
\bibitem [{Note1()}]{Note1}%
  \BibitemOpen
  \bibinfo {note} {By an unfortunate coincidence the strepatividin-coated
  particles are labeled in \cite {Ghofraniha2009} with L (for large) and the
  biotin-coated particles with S (for small). We made the option of keeping our
  notation for particles and linkers. Please be aware of this when contrasting
  our results to those of \cite {Ghofraniha2009}}\BibitemShut {NoStop}%
\bibitem [{\citenamefont {Howard}\ \emph {et~al.}(2019)\citenamefont {Howard},
  \citenamefont {Jadrich}, \citenamefont {Lindquist}, \citenamefont {Khabaz},
  \citenamefont {Bonnecaze}, \citenamefont {Milliron},\ and\ \citenamefont
  {Truskett}}]{Howard2019}%
  \BibitemOpen
  \bibfield  {author} {\bibinfo {author} {\bibfnamefont {M.~P.}\ \bibnamefont
  {Howard}}, \bibinfo {author} {\bibfnamefont {R.~B.}\ \bibnamefont {Jadrich}},
  \bibinfo {author} {\bibfnamefont {B.~A.}\ \bibnamefont {Lindquist}}, \bibinfo
  {author} {\bibfnamefont {F.}~\bibnamefont {Khabaz}}, \bibinfo {author}
  {\bibfnamefont {R.~T.}\ \bibnamefont {Bonnecaze}}, \bibinfo {author}
  {\bibfnamefont {D.~J.}\ \bibnamefont {Milliron}}, \ and\ \bibinfo {author}
  {\bibfnamefont {T.~M.}\ \bibnamefont {Truskett}},\ }\href@noop {} {\bibfield
  {journal} {\bibinfo  {journal} {J. Chem. Phys.}\ }\textbf {\bibinfo {volume}
  {151}},\ \bibinfo {pages} {124901} (\bibinfo {year} {2019})}\BibitemShut
  {NoStop}%
\end{thebibliography}%
%\bibitem{redner} Pavel L. Krapivsky , Sidney Redner and  Eli Ben-Naim, {\it A Kinetic View of Statistical Physics}, Cambridge University Press, New York, 2010.
%\end{thebibliography}

\end{document}

\title{Supplemental Material of ''Smoluchowski equations for linker-mediated irreversible aggregation''}

  \author{J. M. Tavares}
   \email{jmtavares@fc.ul.pt}
    \affiliation{Centro de F\'{i}sica Te\'{o}rica e Computacional, Universidade de Lisboa, 1749-016 Lisboa, Portugal}
    \affiliation{Instituto Superior de Engenharia de Lisboa, ISEL, Avenida Conselheiro Em\'{i}dio Navarro, 1  1950-062 Lisboa, Portugal}

  \author{G. C. Antunes}
   \email{antunes@is.mpg.de}
    \affiliation{Departamento de F\'{\i}sica, Faculdade de Ci\^{e}ncias, Universidade de Lisboa, 
    1749-016 Lisboa, Portugal}
   \affiliation{Centro de F\'{i}sica Te\'{o}rica e Computacional, Universidade de Lisboa, 1749-016 Lisboa, Portugal}
    \affiliation{Max Planck Institute for Intelligent Systems, Stuttgart, Germany}
    \affiliation{Institute for Theoretical Physics IV, University of Stuttgart, Pfaffenwaldring 57, 70569 Stuttgart, Germany}

  \author{C. S. Dias}
   \email{csdias@fc.ul.pt}
    \affiliation{Departamento de F\'{\i}sica, Faculdade de Ci\^{e}ncias, Universidade de Lisboa,     1749-016 Lisboa, Portugal}
    \affiliation{Centro de F\'{i}sica Te\'{o}rica e Computacional, Universidade de Lisboa, 1749-016 Lisboa, Portugal}

  \author{M. M. Telo da Gama}
   \email{mmgama@fc.ul.pt}
    \affiliation{Departamento de F\'{\i}sica, Faculdade de Ci\^{e}ncias, Universidade de Lisboa,     1749-016 Lisboa, Portugal}
    \affiliation{Centro de F\'{i}sica Te\'{o}rica e Computacional, Universidade de Lisboa, 1749-016 Lisboa, Portugal}

  \author{N. A. M. Ara\'ujo}
   \email{nmaraujo@fc.ul.pt}
    \affiliation{Departamento de F\'{\i}sica, Faculdade de Ci\^{e}ncias, Universidade de Lisboa,    1749-016 Lisboa, Portugal}
    \affiliation{Centro de F\'{i}sica Te\'{o}rica e Computacional, Universidade de Lisboa, 1749-016 Lisboa, Portugal}

    \maketitle    
    
\section{Approximation for the kernel}   
Here we briefly review how a Brownian kernel
is approximated by a constant kernel \cite{Chandrasekhar43,Krapivsky2011}. 
The kernels used in this work for the two types of aggregation can be written as $K(i_1,j_1;i_2,j_2)=K_B(i_1,j_1;i_2,j_2) p(i_1,j_1;i_2,j_2)$ and
$K_0(i,j)=K_{B,0}(i,j) p_0(i,j)$,, with the Brownian kernels given by
\begin{equation}
\label{kernel1}
K_B(i_1,j_2;i_2,j_2)=4\pi\left(D_{i_1j_1}+D_{i_2j_2}\right) \frac{R_{i_1j_1}+R_{i_2j_2}}{2},
\end{equation}
and
\begin{equation}
\label{kernel0}
K_{B,0}(i,j)=4\pi\left(D_{ij}+D_{L}\right) \frac{R_{ij}+R_{L}}{2}.
\end{equation}
Here $D_{ij}$, $R_{ij}$ are the diffusion coefficient and the size of a cluster $(i,j)$, respectively; and $D_L$, $R_L$ are the diffusion coefficient and the size of linkers, respectively.
The constant kernel is obtained by assuming that
$D_{ij}=D_P$ and $R_{ij}=R_P$, i.e., all clusters have the same diffusion coefficient and the same size of one single particle. Within this approximation the kernels (\ref{kernel1},\ref{kernel0}) become,
\begin{equation}
\label{kernel1ap}
K_B(i_1,j_2;i_2,j_2)=2 \kappa_P 
\end{equation}
and
\begin{equation}
\label{kernel0ap}
K_{B,0}(i,j)=2\kappa_P \alpha,
\end{equation}
with $\kappa_P=4\pi D_PR_P$ and 
\begin{equation}
\label{alpha}
\alpha=\frac{1}{4}\left(1+\frac{D_L}{D_P}\right) \left(1+\frac{R_L}{R_P}\right).
\end{equation}
Notice that in real systems, according to Stokes-Einstein relation, $R_L/R_P=D_P/D_L=1/\Delta$ and thus $\alpha=(1+\Delta)(1+1/\Delta)/4$. 

 \section{Finite-size effects}
\begin{figure} [thb]
\includegraphics[width=\columnwidth]{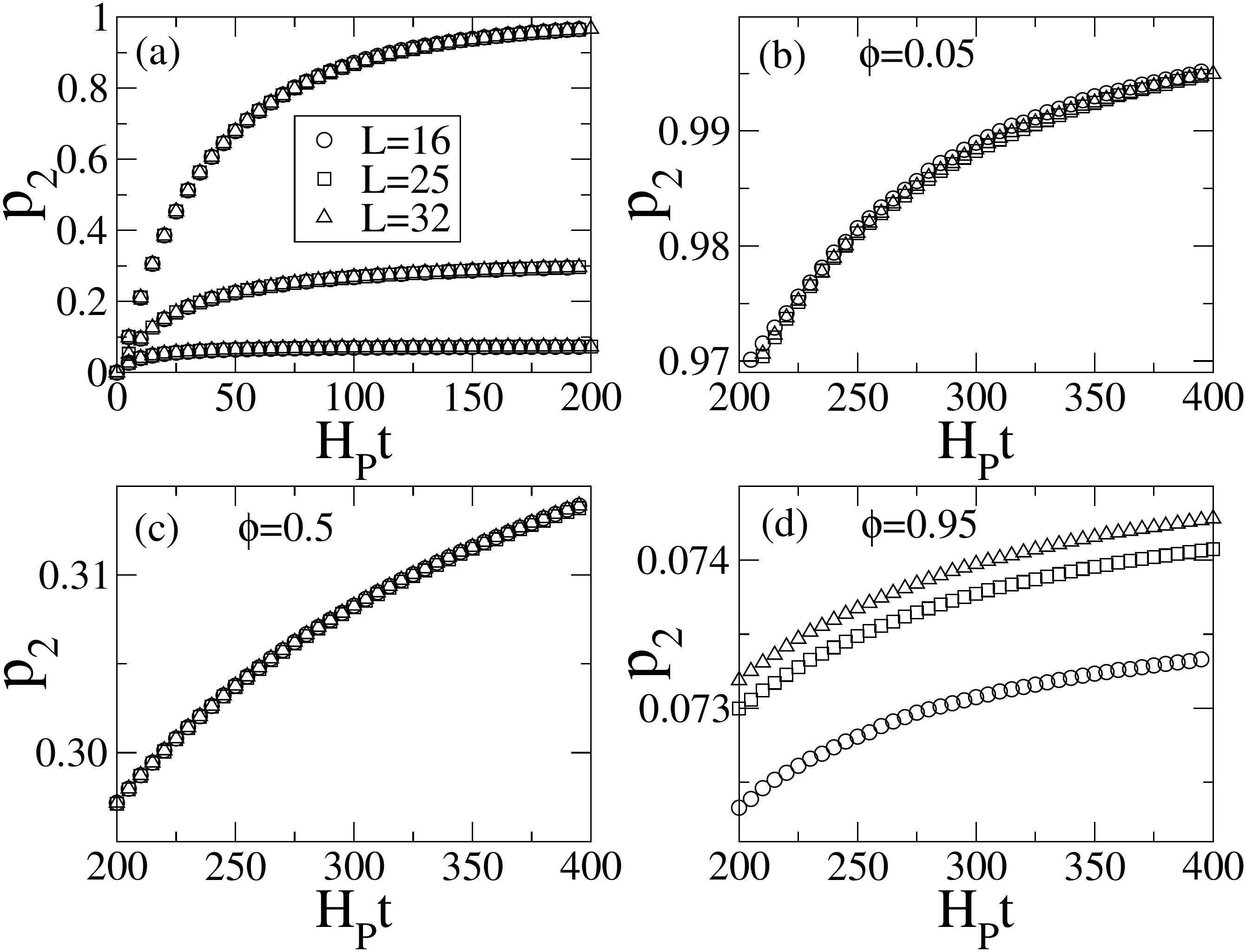}
\caption{Analysis of finite-size effects. Results for the probability $p_2$ obtained with $\Delta=10$ in simulations with $N_{\rm{latt}}=L^3$ sites, for $L=16,25$ and $32$. (a) short times and $\phi=0.05,0.5$ and $0.95$; (b) long times, $\phi=0.05$; (c) long times, $\phi=0.5$; (d) long times, $\phi=0.95$. ~\label{smfig1}}
\end{figure}

 The simulations in the main text were performed 
 on a lattice of size $N_{\rm{latt}}=25^3$. In Fig. \ref{smfig1}, the results obtained for $p_2$, with
  $\Delta=10$ and  $\phi=0.05,0.5$ and $0.95$, on lattices of sizes  $N_{\rm{latt}}=L^3$, with $L=16,25$ and $32$ show that finite-size effects are negligible. All differences found    are less than 1$\%$, even in the worst case  $\phi=0.95$.

\section{Asymptotic results for other $\Delta$}

In the main text,  the  asymptotic values of the probabilities  for $\phi<1-1/f$  are  obtained with $\Delta=10$ (simulation results). For $\Delta=1, 5$ and $20$ we obtained exactly  the same asymptotic values for the probabilities: for $\phi=0.05$, $p_2=1$; for $\phi=0.5$, $p_1=0.6672$ and $p_2=0.3328$. The latter are to be compared with the results of the  theory for $\phi=0.5$: $p_2=1/3$ and $p_1=2/3$.

\section{Comparison to the case without linkers} 
\begin{figure} [thb]
\includegraphics[width=\columnwidth]{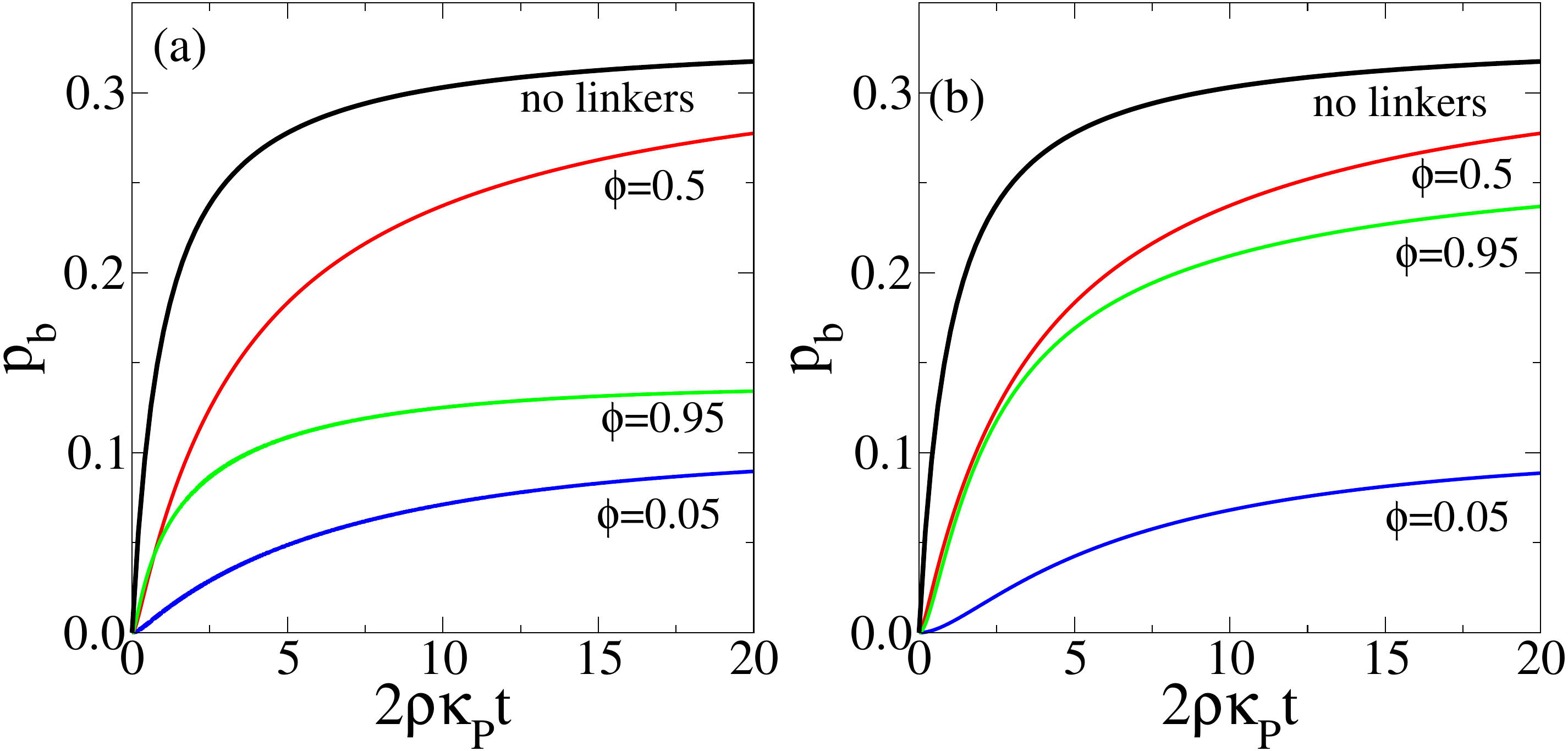}
\caption{Bonding probability as a function of time for a system with no linkers where particles bond directly and for systems with particle-linker aggregation. (a) $\Delta=1$, (b) $\Delta=10$. ~\label{smfig2}}
\end{figure}

       The results obtained in the main text can be compared to those of a system with particles (no linkers) that may bond directly to each other through their $f$ bonding sites. In such a case, the theory with a constant kernel predicts that $p_b=\frac{2}{f} \frac{t}{1+t}$ \cite{Krapivsky2011}. Figure \ref{smfig2} compares this result to those obtained with the approximate theory for particles and linkers for several values of $\Delta$ and $\phi$. At the early stages, the growth of $p_b$ (and therefore of clusters) is much smaller when linkers are present. In the case $\phi=0.5$,  the asymptotic state is the same with or without linkers ($p_b=2/f$).  It is, however, immediately obvious that the introduction of linkers delays cluster formation even at long times. Therefore, the introduction of linkers allows for the tunable slowing down of aggregation, and in particular enables the study and exploitation of early-stage dynamics.
    %\begin{thebibliography}*
%\bibliography{library}
\bibliography{linkers}
%\bibitem{redner} Pavel L. Krapivsky , Sidney Redner and  Eli Ben-Naim, {\it A Kinetic View of Statistical Physics}, Cambridge University Press, New York, 2010.
%\end{thebibliography}